\documentclass[12pt]{article}
\usepackage{amssymb}
\usepackage{graphics}

\catcode`\@=11
\def\@citex[#1]#2{%
\if@filesw \immediate \write \@auxout {\string \citation {#2}}\fi
\@tempcntb\m@ne \let\@h@ld\relax \def\@citea{}%
\@cite{%
  \@for \@citeb:=#2\do {%
    \@ifundefined {b@\@citeb}%
      {\@h@ld\@citea\@tempcntb\m@ne{\bf ?}%
      \@warning {Citation `\@citeb ' on page \thepage \space undefined}}%
      {\@tempcnta\@tempcntb \advance\@tempcnta\@ne%
      \@tempcntb\number\csname b@\@citeb \endcsname \relax%
      \ifnum\@tempcnta=\@tempcntb 
        \ifx\@h@ld\relax%
          \edef \@h@ld{\@citea\csname b@\@citeb\endcsname}%
        \else%
          \edef\@h@ld{\ifmmode{-}\else--\fi\csname b@\@citeb\endcsname}%
        \fi%
      \else
        \@h@ld\@citea\csname b@\@citeb \endcsname%
        \let\@h@ld\relax%
      \fi}%
    \def\@citea{,\penalty\@highpenalty\,}%
  }\@h@ld
}{#1}}

\def\@citeb#1#2{{[#1]\if@tempswa , #2\fi}}
\def\@citeu#1#2{{$^{#1}$\if@tempswa , #2\fi }}
\def\@citep#1#2{{#1\if@tempswa , #2\fi}}

\def\bcites{         
        \catcode`\@=11
        \let\@cite=\@citeb
        \catcode`\@=12
}

\def\upcites{         
        \catcode`\@=11
        \let\@cite=\@citeu
        \catcode`\@=12
}

\def\plaincites{      
        \catcode`\@=11
        \let\@cite=\@citep
        \catcode`\@=12
}

\newcount\hour
\newcount\minute
\newtoks\amorpm
\hour=\time\divide\hour by 60
\minute=\time{\multiply\hour by 60 \global\advance\minute by-\hour}
\edef\standardtime{{\ifnum\hour<12 \global\amorpm={am}%
        \else\global\amorpm={pm}\advance\hour by-12 \fi
        \ifnum\hour=0 \hour=12 \fi
        \number\hour:\ifnum\minute<10 0\fi\number\minute\the\amorpm}}
\edef\militarytime{\number\hour:\ifnum\minute<10 0\fi\number\minute}

\def\draftlabel#1{{\@bsphack\if@filesw {\let\thepage\relax
   \xdef\@gtempa{\write\@auxout{\string
      \newlabel{#1}{{\@currentlabel}{\thepage}}}}}\@gtempa
   \if@nobreak \ifvmode\nobreak\fi\fi\fi\@esphack}
        \gdef\@eqnlabel{#1}}
\def\@eqnlabel{}
\def\@vacuum{}
\def\marginnote#1{}
\def\draftmarginnote#1{\marginpar{\raggedright\scriptsize\tt#1}}
\def\draft{
        \pagestyle{plain}
        \overfullrule=2pt
        \oddsidemargin -.5truein
        \def\@oddhead{\sl \phantom{\today\quad\militarytime} \hfil
        \smash{\Large\sl DRAFT} \hfil \today\quad\militarytime}
        \let\@evenhead\@oddhead
        \let\label=\draftlabel
        \let\marginnote=\draftmarginnote
        \def\ps@empty{\let\@mkboth\@gobbletwo
        \def\@oddfoot{\hfil \smash{\Large\sl DRAFT} \hfil}
        \let\@evenfoot\@oddhead}
        \def\@eqnnum{(\theequation)\rlap{\kern\marginparsep\tt\@eqnlabel}%
        \global\let\@eqnlabel\@vacuum}  }

\def\section{\@startsection {section}{1}{\z@}{3.ex plus 1ex minus
 .2ex}{2.ex plus .2ex}{\large\bf}}
\def\subsection{\@startsection{subsection}{2}{\z@}{2.75ex plus 1ex minus
 .2ex}{1.5ex plus .2ex}{\bf}}        

\def\appendix{{\newpage\section*{Appendix}}\let\appendix\section%
        {\setcounter{section}{0}
        \gdef\thesection{\Alph{section}}}\section}

\def\abstract{\if@twocolumn
\section*{Abstract}
\else 
\begin{center}
{\bf Abstract\vspace{-.5em}\vspace{0pt}}
\end{center}
\quotation
\fi}

\catcode`\@=12

\newcommand{\beq}{\begin{equation}}
\newcommand{\eeq}{\end{equation}}
\newcommand{\beqa}{\begin{eqnarray}}
\newcommand{\eeqa}{\end{eqnarray}}
\newcommand{\dd}{{\rm d}}

\newcommand{\Z}{{\bf Z}}
\newcommand{\ZZ}{{\mathbb Z}}

\newcommand{\RR}{{\mathbb R}}

\newcommand{\CC}{{\mathbb C}}
\newcommand{\PP}{{\mathbb P}}
\newcommand{\e}{\,{\rm e}}
\newcommand{\CP}{{\CC\PP}}

\newcommand{\ds}{\displaystyle}
\newcommand{\ts}{\textstyle}

\newcommand{\be}{\begin{equation}}
\newcommand{\ee}{\end{equation}}
\newcommand{\bea}{\begin{eqnarray}}
\newcommand{\eea}{\end{eqnarray}}

\def\to{\rightarrow}

\def\lae{\mathrel{\mathop{\smash{\lower .5 ex \hbox{$\stackrel<\sim$}}}}}
\def\lae{\mathrel{\mathop{\smash{\lower .5 ex \hbox{$\stackrel>\sim$}}}}}

\def\l:{\mathopen{:}\,}
\def\r:{\,\mathclose{:}}

\catcode`\@=11
\def\theequation{\arabic{equation}}
\catcode`\@=12

\bcites

\catcode`\@=11
\def\theequation{\thesection.\arabic{equation}}
\@addtoreset{equation}{section}
\@addtoreset{footnote}{section}
\@addtoreset{footnote}{subsection}
\catcode`\@=12

\newcommand{\eps}{\epsilon}
\newcommand{\opsi}{\overline{\psi}}
\newcommand{\oQ}{\overline{Q}}
\newcommand{\btheta}{\overline{\theta}}

\newcommand{\bPhi}{\overline{\Phi}}
\newcommand{\bP}{\overline{P}}
\newcommand{\bp}{\overline{p}}
\newcommand{\bphi}{\overline{\phi}}
\newcommand{\bpsi}{\overline{\psi}}

\newcommand{\bm}{\overline{m}}

\newcommand{\bchi}{\overline{\chi}}
\newcommand{\blambda}{\overline{\lambda}}
\newcommand{\bsigma}{\overline{\sigma}}
\newcommand{\bSigma}{\overline{\Sigma}}

\newcommand{\bD}{\overline{D}}

\newcommand{\cD}{{\cal D}}
\newcommand{\rp}{{{\rm Re}\,p}}
\newcommand{\vp}{\varphi_P}
\newcommand{\cJ}{{\cal J}}
\newcommand{\rre}{{\rm Re}}
\newcommand{\iim}{{\rm Im}}
\newcommand{\ra}{\rightarrow}
\newcommand{\wh}{\widehat}

\newcommand{\tY}{{\tilde Y}}
\newcommand{\bY}{{\overline Y}}
\newcommand{\tL}{\tilde{L}}
\newcommand{\tG}{\tilde{G}}
\newcommand{\tJ}{\tilde{J}}

\newcommand{\tm}{\tilde{m}}
\newcommand{\tpsi}{\tilde{\psi}}

\begin{document}

\begin{titlepage}

\begin{center}

\today\hfill
hep-th/0104202\\
\hfill HUTP-01/A019

\vskip 1.5 cm
{\large \bf
Duality of the Fermionic 2d Black Hole and
${\cal N}=2$ Liouville\\[0.1cm]
 Theory as Mirror Symmetry}
\vskip 1 cm 
{~~~\,Kentaro Hori~~~~~~~~~\, and \,~~~~~~~~Anton Kapustin~\,}\\
\vskip 0.5cm
{\it Jefferson Physical Laboratory
~~~~~~~School of Natural Sciences~~~\\
~~~~~~~\,Harvard University~~~~~~~~~~~
Institute for Advanced Study~\,\\
Cambridge, MA 02138, U.S.A.~~~\,\,
Princeton, NJ 08540, U.S.A.~}\\

\end{center}

\vskip 0.5 cm
\begin{abstract}

We prove the equivalence of the $SL(2,\RR)/U(1)$ Kazama-Suzuki model,
which is a fermionic generalization of the 2d Black Hole,
and ${\cal N}=2$ Liouville theory.
We show that this duality is an example of mirror symmetry.
The essential part of the derivation is
to realize the fermionic 2d Black Hole as the low energy limit
of a gauged linear sigma-model.
Liouville theory is obtained by dualizing
the charged scalar fields and
taking into account the vortex-instanton effects,
as proposed recently in non-dilatonic models. The gauged linear 
sigma-model we study has many useful generalizations which we
briefly discuss. In particular,
we show how to construct a variety of dilatonic superstring backgrounds
which generalize the fermionic 2d Black Hole and admit a mirror description
in terms of Toda-like theories.

\end{abstract}

\end{titlepage}

\newpage

\section{Introduction}\label{sec:intro}

The mirror dual of an ${\cal N}=2$
supersymmetric non-linear sigma-model on a toric variety has been derived
in \cite{HV} by realizing the model
as the low energy limit of a gauged linear sigma-model \cite{phases},
and dualizing the phases of charged scalar fields.
This can be viewed as T-duality applied to
the fibers of a torus fibration.
When a circle fiber shrinks to zero size at some locus of the base, 
one could na\"\i vely expect that the dual circle blows up
at the same locus. What really happens is the following.
To each such degenerating fiber there corresponds a superpotential term,
generated by the vortex-instanton of the gauge system
(analogously to \cite{Polyakov}),
that diverges toward the degeneration locus.
The superpotential also breaks the rotational symmetry of the dual theory,
accounting for the loss of winding number in the original system
due to the degeneration of the circle.
This is the story for $(2,2)$ supersymmetric non-dilatonic
sigma-models on toric manifolds,
but it would be interesting to see how universal this phenomenon is.

Some time ago, Fateev, Zamolodchikov and Zamolodchikov (FZZ) \cite{FZZ}
conjectured a duality between
the conformal field theory of a two-dimensional Euclidean black hole
\cite{2dbh} and a Landau-Ginzburg theory, called {\it sine-Liouville theory}.
The 2d Black Hole is defined as the level $k$
$SL(2,\RR)/U(1)$ coset model and 
has the following target-space metric and dilaton for large $k$
\beq
\begin{array}{l}
\ds \dd s^2=k[\dd \rho^2+\tanh^2\!\!\rho\,\dd\varphi^2],\\
\Phi=\Phi_0-2\log\cosh \rho.
\end{array}
\label{intro:cigar}
\eeq
Here $\varphi$ is a periodic variable of period $2\pi$. The coset theory
is well-defined for all $k>2$.
On the other hand, the sine-Liouville theory is
a theory of scalar fields $-\infty<\varrho<\infty$ and
$\vartheta\equiv\vartheta+2\pi$ with the following action
\beq
\widetilde{S}={1\over 4\pi}\int \left[
{1\over k-2}(\dd \varrho)^2+{1\over k}(\dd\vartheta)^2
-{1\over k-2}R_h\varrho
+\mu^2\e^{-\varrho}\cos\vartheta\right]\sqrt{h}\dd^2x,
\eeq
where $h$ is the world-sheet metric (with Ricci scalar $R_h$)
and $\mu$ is some mass scale. 
We refer the reader to \cite{KKK} for a review of this conjectural duality.
The duality was used in \cite{KKK} as the starting point for the
Matrix Model formulation of string theory
in the black hole background.

The 2d Black Hole has an asymptotic region, $\rho\ra +\infty$,
where the geometry is that of a cylinder of radius $\sqrt{k}$
and the dilaton is linear, $\Phi\sim -2\rho$.
At $\rho=0$ the circle shrinks to zero size,
and therefore the overall geometry is that of a semi-infinite cigar.
Sine-Liouville theory also has an asymptotic region, $\varrho\ra \infty$,
where the potential is exponentially small
and the theory is the sigma-model on a cylinder of radius
$1/\sqrt{k}$ with a linear dilaton
$\widehat{\Phi}\sim -\varrho/(k-2)$. Note that the sine-Liouville potential
is unbounded from below, and therefore for small $k$, where the
radius of the cylinder is large and semiclassical reasoning is valid, we
expect the model to be unstable. This corresponds to the fact that
the coset model is well-defined only for $k>2$.

If we compare the radii of the two asymptotic regions,
we notice that the two theories may be related by T-duality.
The shrinking of the circle as one goes towards $\rho=0$ on 
the 2d Black Hole side
corresponds to the exponentially growing potential which breaks
rotational symmetry on the sine-Liouville side.
Thus FZZ duality is strongly reminiscent of 
mirror duality between $(2,2)$ sigma-models and $(2,2)$ Landau-Ginzburg
models mentioned above.

In this paper, we prove the supersymmetric version of FZZ duality
using the method of \cite{HV}. 
Instead of a 2d Black Hole we consider a fermionic 2d Black Hole,
defined as the level $k$ $SL(2,\RR)/U(1)$
Kazama-Suzuki supercoset model \cite{KazS}, and instead of sine-Lioville
theory we consider ${\cal N}=2$ supersymmetric Liouville theory~\cite{IvKriv}. 
This duality was conjectured in \cite{GK1} from
the space-time point of view;
closely related ideas were discussed earlier in~\cite{MV,GM,OV},
and the duality was studied more recently in \cite{ES1}. 
The supercoset model can be viewed as an ${\cal N}=1$ supersymmetric
sigma-model with target-space metric~(\ref{intro:cigar}).
The action for ${\cal N}=2$ Liouville theory on a flat world-sheet is given by
\beq
\widetilde{S}={1\over 2\pi}\int \dd^2x\left[\,
\int \dd^4\theta {1\over 2k}|Y|^2
+{1\over 2}\left(\int\dd^2\theta
\,\mu\e^{-Y}+h.c.\right)\,\right],
\label{intro:Liouville}
\eeq
where $Y$ is a chiral superfield with period $2\pi i$ and $\mu$ is a mass
scale. (A linear dilaton is hidden in this action.)
As in the bosonic case,
the two theories have asymptotic regions that are
related by T-duality, and the shrinking of the circle on one side
corresponds to growing superpotential breaking rotational
symmetry on the other side. Unlike in the bosonic case, the supercoset
theory is well-defined for all $k>0$. This corresponds to the fact
that ${\cal N}=2$ Liouville theory makes sense for all $k>0$.

The crucial part of our proof is showing that the
$(2,2)$ superconformal field theory of the fermionic 2d Black Hole
arises as the infrared limit of a certain super-renormalizable gauge theory.
The candidate system is the $U(1)$ gauge theory with two chiral superfields
$\Phi$ and $P$ on which the gauge transformation acts as
$\Phi\to\e^{i\alpha}\Phi$ and $P\to P+i\alpha$.
The action is
\beq
S={1\over 2\pi}\int \dd^2x\,\dd^4\theta
\left[\,
\bPhi \e^V\Phi+{k\over 4}(P+\bP+V)^2-{1\over 2e^2}|\Sigma|^2
\,\right].
\label{intro:action}
\eeq
We will first give some numerical evidence. We will show that
the sigma-model that arises after integrating out the
gauge multiplet flows under one-loop renormalization group
flow to the supersymmetric sigma-model with target-space metric
(\ref{intro:cigar}).
We will explicitly see how the linear dilaton in the asymptotic region is
generated.
The one-loop approximation is valid for large $k$.
To go beyond this approximation, we
compute the infrared central charge of the above gauged linear sigma-model (GLSM).
Following \cite{minimal,SW},
we identify the right-moving ${\cal N}=2$ superconformal algebra
in the ring of left-chiral operators.
The classical gauge theory (\ref{intro:action}) has both vector and axial 
R-symmetries, but on the quantum level
the axial R-symmetry is anomalously broken.
However, one can modify the current using the
field $P$ to make it conserved.
This allows us to identify the right-moving R-current,
and then the full ${\cal N}=2$ superconformal algebra.
The correction terms in the superconformal currents
are linear in $P$ and generate linear dilaton in the asymptotic region.
(Alternatively, one can obtain the whole current superfield
by cancelling the Konishi anomaly~\cite{Konishi} associated to the axial 
anomaly.)
We find that the central charge is
\beq
c=3\left(1+\frac{2}{k}\right),
\label{intro:c}
\eeq
which coincides with the central charge of the level $k$ $SL(2,\RR)/U(1)$
Kazama-Suzuki model. The asymptotic behavior of the target-space metric
also agrees in the two theories.
Then we argue the uniqueness of the SCFT with this value of the central charge,
asymptotic behaviour, and symmetries.
This establishes that our gauge theory (\ref{intro:action})
flows to the fermionic 2d Black Hole for all $k>0$.

Since the UV central charge of the GLSM
is 9, and the IR central charge~(\ref{intro:c}) becomes arbitrarily large
as $k\ra 0$, one may wonder how these results are consistent with 
Zamolodchikov's $c$-theorem~\cite{ctheorem}. The resolution of this
apparent paradox is well known~\cite{Seibergnotes,KSdgf}.
Technically, the $c$-theorem is not applicable here because the
IR conformal field theory violates one of the assumptions made
in~\cite{ctheorem}, namely the assumption that there exists a normalizable
$SL(2,\CC)$-invariant vacuum state. A more satisfactory explanation is that
in general the central charge is not a good measure of the number of degrees
of freedom. For example, if one does not assume normalizability of vacuum,
Cardy's formula~\cite{Cardy} says that the 
growth of the density of states is 
determined not by $c$, but by $c_{eff}=c-24 h_{min}$, where $h_{min}$ 
is the lower boundary of the spectrum of $L_0$.\footnote{We assume that world-sheet
parity is a symmetry of the theory. Otherwise $h_{min}$ is defined as the smaller
of the lower boundaries of the spectra of $L_0$ and $\tL_0$.}
If a unitary CFT has a 
normalizable vacuum, then $h_{min}=0$, but in general the effective number of 
degrees
of freedom is different from $c$. For the supercoset model $h_{min}=\frac{1}{4k}$
(this can be derived either by using the fact that the supercoset is asymptotic
to a linear dilaton theory with background charge $Q=1/{\sqrt k}$ and applying
the Seiberg bound~\cite{Seibergnotes}, or
by the direct analysis of the operator spectrum),
and therefore $c_{eff}=3$. Thus the effective
number of degrees of freedom decreases as one flows towards the infrared,
in agreement with expectations.

Once the flow to the fermionic 2d Black Hole is established,
the rest is a straightforward generalization of~\cite{HV}.
Dualizing the phase of $\Phi$ and the imaginary part of $P$,
we obtain twisted chiral superfields $Y$ and $Y_P$
of period $2\pi i$.
The superpotential of the dual system is
\beq
\widetilde{W}=\Sigma (Y+Y_P)+\mu\e^{-Y},
\eeq
where the term linear in $\Sigma$ is present already at the classical level,
and the exponential term is generated by the vortex of $\Phi$.
Note that the $P$-vortex is absent, and therefore no nonperturbative
superpotential is generated for $Y_P$.
The K\"ahler potential is
\beq\label{kahlerint}
K=-{1\over 2e^2}|\Sigma|^2-{1\over 2k}|Y_P|^2+\ldots,
\eeq
where dots denote a possible correction term that vanishes in the
asymptotic region ${\rm Re}\, Y\to\infty$.
In the infrared limit $e\to\infty$ it is appropriate to integrate out
$\Sigma,$ and this gives a constraint $Y+Y_P=0$.
Thus, we obtain a theory of a single periodic chiral superfield $Y$
with the superpotential $\e^{-Y}$. Using the uniqueness
of the supersymmetric coset, one can show that the corrections to the
K\"ahler potential indicated by dots in~(\ref{kahlerint}) are in fact
absent.  Note that in general the methods of~\cite{HV} do not
allow to control the K\"ahler potential. What makes the present case
different is that one can continuously deform the gauge 
theory~(\ref{intro:action}) to the ${\cal N}=2$ Liouville theory
without breaking any symmetries. Since the supersymmetric coset is rigid,
this implies that the infrared limit of the theory~(\ref{intro:action})
is equivalent to ${\cal N}=2$ Liouville theory. This alternative way of 
deriving the mirror dual is less general
than that used in~\cite{HV}, but provides more information about the dual 
theory.

We also describe some obvious generalizations of the 
model~(\ref{intro:action}), compute their infrared central charge and
find mirror duals. Some of these models flow to non-trivial $(2,2)$
superconformal field theories and can be used to construct
a variety of higher-dimensional superstring backgrounds with a non-constant
dilaton and fermionic symmetries. Others are massive field theories
which upon integrating out the gauge fields reduce to sigma-models
on ``squashed'' toric varieties.
Mirror symmetry relates these sigma-models
to Landau-Ginzburg models; for example, the sigma-model on a ``squashed''
$\CC{\mathbb P}^1$ (the supersymmetric ``sausage model'')
is mirror to the ${\cal N}=2$ sine-Gordon model with a finite K\"ahler
potential. In fact, in this particular case both theories are integrable,
and their equivalence has been conjectured by Fendley and 
Intriligator~\cite{FIpr}.
(The squashed toric sigma models and the mirrors
are also introduced and studied from a different but related
point of view in \cite{AHKT}.)

\section{The Gauged Linear Sigma-Model}
The field content of the gauged linear sigma-model will be
the following: two chiral superfields $\Phi$ and $P$ and
a vector superfield $V$. Our superfield conventions are collected in
Appendix A. The gauge transformations laws are
defined to be
\beqa
&&\Phi\to\e^{i\Lambda}\Phi,
\nonumber\\
&&P\to P+i\Lambda,
\nonumber\\
&&V\to V-i\Lambda+i\overline{\Lambda},
\label{gauge}
\eeqa
where $\Lambda$ is a chiral superfield, 
$\bD_+\Lambda=\bD_-\Lambda=0.$ We take the gauge group to be
$U(1)$, and $\iim\, P$ is periodically identified with period $2\pi$.

The action of the system is
\beq
S={1\over 2\pi}\int\dd^2x\,\dd^4\theta\left[
\bPhi \e^V\Phi+{k\over 4}(P+\bP+V)^2-{1\over 2e^2}|\Sigma|^2\right].
\label{action}
\eeq
Here $\Sigma=\bD_+ D_-V$ is a twisted chiral superfield,
$\bD_+\Sigma=D_-\Sigma=0$.
We did not include the Fayet-Iliopoulos term as it can be absorbed into $P$.
Neither did we include its superpartner, the theta-angle, since it breaks
world-sheet parity, while we want the theory to flow to a parity-invariant
supercoset model (see Appendix D for details about the definition
of world-sheet parity for the coset models).

The chiral superfield $P$ can be gauged away completely, after
which one is left with $\Phi$ and a massive vector superfield
described by $V$. Thus the action~(\ref{action}) describes 
massive ${\cal N}=2$ QED.
Alternatively, one can choose the Wess-Zumino gauge for 
$V$ and retain $P$. Then the action in terms of component fields reads
\beqa
&&\!\!\!\!{1\over 2\pi}\int\dd^2x~\Biggl[
-\cD^{\mu}\bphi\cD_{\mu}\phi
+i\bpsi_-(\cD_0+\cD_1)\psi_-
+i\bpsi_+(\cD_0-\cD_1)\psi_+
+D|\phi|^2+|F|^2~~~~~~~~~~~
\nonumber\\[-0.1cm]
&&~~~~~~~~~~~~~~~
-|\sigma|^2|\phi|^2
-\bpsi_-\sigma\psi_+
-\bpsi_+\bsigma\psi_-
-i\bphi\lambda_-\psi_+
+i\bphi\lambda_+\psi_-
+i\bpsi_+\blambda_-\phi
-i\bpsi_-\blambda_+\phi
\nonumber\\[0.1cm]
&&~~~~~~~~~~~~
+{k\over 2}\Bigl(-\cD^{\mu}\bp\cD_{\mu}p
+i\bchi_-(\partial_0+\partial_1)\chi_-
+i\bchi_+(\partial_0-\partial_1)\chi_+
+D(p+\bp)+|F_P|^2
\nonumber\\
&&~~~~~~~~~~~~~~~
-|\sigma|^2
+i\chi_+\lambda_-
-i\chi_-\lambda_+
+i\bchi_+\blambda_-
-i\bchi_-\blambda_+\Bigr)
\nonumber\\
&&~~~~~~~~~~~~
+{1\over 2e^2}\Bigl(
-\partial^{\mu}\bsigma\partial_{\mu}\sigma
+i\blambda_-(\partial_0+\partial_1)\lambda_-
+i\blambda_+(\partial_0-\partial_1)\lambda_+
+v_{01}^2+D^2\Bigr)\Biggr].
\label{actioncomp}
\eeqa
Here $\phi$ and $p$ are the lowest components of $\Phi$ and $P$,
respectively, $\psi$ and $\chi$ are their superpartners, and 
$v_{\mu},\lambda,$
and $D$ are components of a vector multiplet in the Wess-Zumino gauge.
$\cD_{\mu}\phi$ and $\cD_{\mu}\psi_{\pm}$ are the standard covariant
derivatives, while $\cD_{\mu}p:=\partial_{\mu}p+iv_{\mu}$.
After one gauges away the imaginary part of $p$, one can see that the
gauge field and its superpartners have mass $e\sqrt k$.

This field theory is free in the UV and super-renormalizable. We are 
interested in its infrared limit. At energies
much lower than $e\sqrt k$ one can integrate out $\Sigma$ and set
the D-term potential to zero. The D-term is given by
$$
D(\phi,p)=|\phi|^2+k\,\rre\, p.
$$
To obtain the low-energy
effective action for $\Phi$ we set $\iim\, p=0$ (this is a gauge
choice), express $\rre\, p$ in terms of $\phi$ by means of 
$D(\phi,p)=0$,
and integrate out $V$ omitting the last term in the action
(because the infrared limit is equivalent to taking $e\ra\infty$).
Equivalently, we can take the 
flat space parametrized by $\phi$ and $p$ with K\"ahler
potential 
$$
K(\phi,p)=|\phi|^2+\frac{k}{2}|p|^2,
$$
and compute its K\"ahler quotient with respect to
the action of $U(1)$ given by
$$
\phi\ra \phi e^{i\lambda}, \quad p\ra p+i\lambda.
$$

Either way, one concludes that the low-energy theory is
described by a supersymmetric non-linear sigma-model with 
the following target space metric:
\begin{equation}\label{ourcoset}
ds^2=(1+\frac{r^2}{k})dr^2+\frac{r^2}{1+\frac{r^2}{k}}d\theta^2.
\end{equation}
Here $r=\sqrt{2}|\phi|\in [0,+\infty),\ \theta=\arg\phi\in \RR/(2\pi\ZZ).$ 
This metric is smooth near the origin $r=0$, while for $r\ra\infty$
it approaches a flat metric on a cylinder of circumference
$2\pi \sqrt{k}.$ Thus it describes a cigar, i.e.
a 2d Riemannian manifold diffeomorphic to $\RR^2$ with a metric 
which has a $U(1)$ isometry and asymptotes to a flat metric on 
a cylinder.

The metric~(\ref{ourcoset}) is different from the usual 2d Black Hole
metric~\cite{2dbh}. If one sets $r=\sqrt{k}\sinh \rho$, 
the metric~(\ref{ourcoset}) takes the form
\begin{equation}\label{ours}
ds^2=k\left(\cosh^4\rho\ d\rho^2+\tanh^2\rho\ d\theta^2\right),
\end{equation}
while the 2d Black Hole metric is
\begin{equation}\label{witten}
ds^2=k\left(d\rho^2+\tanh^2\rho\ d\theta^2\right).
\end{equation}

Qualitatively, the difference between the two metrics is the following.
Let us define a natural
``radial'' variable $v\in[0,+\infty)$ by
\begin{equation}
v(\rho)=\int_0^\rho \sqrt{g_{\rho\rho}(\rho)}d\rho.
\end{equation}
In terms of $v,\theta$ any
cigar-like metric has the form
$$
ds^2=dv^2+F^2(v)\ d\theta^2
$$
for some function $F(v)$ which approaches a constant for $v\ra \infty$.
For our metric the difference $F(v)-\sqrt{k}$ is of order $1/v$ for 
large $v$, while for the 2d Black Hole it is exponentially small.

The metric (\ref{ours}) does not define a conformal
field theory, and flows in a non-trivial way under the renormalization
group. We will show that the end-point of the flow
is the fermionic 2d Black Hole.

Let us conclude this section by listing the symmetries of the 
action~(\ref{action}).
Classically, we have $(2,2)$ supersymmetry,
axial and vector R-symmetries (such that the lowest
components of $\Phi$ and $P$ have zero R-charges), world-sheet parity,
and a global non-R symmetry which which shifts $\iim\, p$ by a constant
and leaves all other fields invariant. The generator
of the latter symmetry will be called momentum, since the corresponding
symmetry in the low energy nonlinear sigma-model shifts $\theta$ and leaves 
$\rho$ and the fermions invariant. Quantum-mechanically, the na{\"\i}ve axial
R-current is anomalous, but one can nevertheless define a conserved 
gauge-invariant axial R-current. This is discussed in detail in 
Section~\ref{ccharge}. In the infrared the R-symmetry gets promoted to
a pair of affine $U(1)$ current algebras 
(one left-moving and one right-moving). In contrast, the left and right 
components of the momentum current are not conserved separately even in the
infrared. Nevertheless, this symmetry will play an important role in our
analysis.

\section{Flow to 2d Black Hole I: One-Loop Approximation}\label{flowone}
For $r\ra\infty$ the metric (\ref{ours}) is flat, and therefore
is unchanged by the RG flow. In other words, the RG flow
deforms the cigar metric without modifying its asymptotic behavior.
We would like to show that in the infrared the supersymmetric sigma-model 
with the metric (\ref{ours}) flows to the fermionic 2d Black Hole (\ref{witten}) 
with the same value of the asymptotic radius.
In this section we limit ourselves to the one-loop approximation, 
which is valid for large enough $k$. 

Consider the one-loop beta-function for the sigma-model metric:
\begin{equation}
\beta_{ij}=-\frac{1}{2\pi}R_{ij}.
\end{equation}
Its only zero is a flat metric, and since any cigar has a nonzero
curvature near the tip, na{\"\i}vely it appears that a cigar-like
metric cannot be a fixed point of the RG flow. The resolution
of this puzzle is well-known~(see e.g.~\cite{Friedan}) and is 
related to the possibility of
having a dilaton gradient. In the usual formulation, the dilaton
affects the coupling of the sigma-model to a curved world-sheet
metric. Alternatively, if one prefers to stay on a flat world-sheet,
one may say that a non-trivial dilaton gradient in space-time
is equivalent to assigning a non-trivial Weyl transformation law
to target-space coordinates. 

Once the possibility of a non-trivial Weyl transformation law
for $X^i$ is recognized, it is easy to see in what sense a
cigar can be invariant under RG flow. Let us fix a conformally-flat
gauge for the space-time metric $G_{ij}$, so that it has the form
\begin{equation}\label{confgauge}
ds^2=e^{\Psi(u)}\left(du^2+d\theta^2\right).
\end{equation}
The function $\Psi(u)$ does not depend on $\theta$ because we are only
interested in the sigma-models which have a $U(1)$ isometry.
The tip of the cigar corresponds to $u\ra -\infty$, while
the cylindrical asymptotics is reached for $u\ra +\infty$.
{}From the known behavior at the tip and at infinity we infer
that
\bea\label{asympt}
\Psi(u)&\sim & 2u + \ldots\quad {\rm for}\ u\ra -\infty,\\
\Psi(u)&\sim & \log k +\ldots\quad {\rm for}\ u\ra +\infty. \nonumber
\eea
The functions $\Psi(u)$ and $F(v)$ are related as follows:
\begin{equation}\label{FPsi}
F(v)=e^{\Psi(u)/2},\quad v=\int_{-\infty}^u e^{\Psi(u)/2} du.
\end{equation}

Note that both (\ref{confgauge}) and (\ref{asympt}) are left 
invariant by reparametrizations $u\ra u+c,
\theta\ra \theta+c'$, where $c,c'$ are constants.
This is what remains of reparametrization invariance after we
fix the gauge (\ref{confgauge},\ref{asympt}).
Hence the most general transformation law for $u$ and $\theta$ 
under the Weyl rescaling of the world-sheet metric by $t^2$ is
$$
u\ra u + a t,\ \theta\ra \theta + a't,
$$
where $a,a'$ are real constants. Saying that the metric approaches
a fixed limit under such a modified Weyl transformation is
equivalent to saying that for $\mu\ra\infty$ the function
$\Psi(u,t)$ depends only on the difference $u - at$:
$$
\Psi(u,t)\ra \Psi_{IR}(u-at).
$$
Since $\Psi$ does not depend on $\theta$, by a $t$-dependent
reparametrization of $\theta$ one can make $a'=0$.

The one-loop RG equation for $\Psi$ is
\begin{equation}\label{oneloopRG}
\frac{\partial \Psi(u,t)}{\partial t}=\frac{1}{4\pi} e^{-\Psi(u,t)}\ 
\frac{\partial^2 \Psi(u,t)}{\partial u^2}.
\end{equation}
Letting $\Psi(u,t)=\Psi_{IR}(u-at),$ we obtain an equation 
for $\Psi_{IR}(u)$:
$$
\frac{1}{4\pi} e^{-\Psi_{IR}(u)}\ 
\Psi_{IR}''(u)+a\Psi_{IR}'(u)=0.
$$
The general solution of this equation is
$$
e^{\Psi_{IR}(u)}=\frac{1}{e^{-\lambda(u-b)}+\frac{4\pi a}{\lambda}},
$$
where $\lambda,b$ are constants.
Imposing the conditions (\ref{asympt}), we obtain 
$$
\lambda=2,\quad a=\frac{1}{2\pi k},\quad e^{\Psi_{IR}(u)}=
\frac{1}{e^{-2(u-b)}+\frac{1}{k}}.
$$
Thus $\Psi_{IR}(u)$ is completely fixed
up to residual reparametrizations of $u$ (shifts by a constant).
In addition, the constant $a$ in the modified Weyl transformation
law is determined by the asymptotic radius of the cigar.
By a change of variables
$$
\sqrt{k}\tanh \rho=e^{\Psi_{IR}(u)/2}
$$
the metric 
$$
ds^2=e^{\Psi_{IR}(u)}(du^2+d\theta^2)
$$
is transformed to the form Eq.~(\ref{witten}).
This proves that the only cigar-like fixed point of the one-loop 
RG equations is the 2d Black Hole.

We now would like to show that our metric (\ref{ours}) indeed 
flows to this infrared fixed point. We set
$$
\Psi(u,t)=f(u - t/(2\pi k),t),
$$
and solve numerically the RG equation for $f(u,t)$. The initial
condition is implicitly given by the metric (\ref{ours}). Explicitly,
$\Psi(u,0)=\Psi_0(u)$ can be written in a parametric form
$$
e^{\Psi_0(u(r))}=\frac{kr^2}{1+r^2},\quad
u(r)=\log r + \frac{r^2}{2}.
$$

It is useful to note that the equation~(\ref{oneloopRG})
is invariant with respect to the transformation
$$
\Psi(u,t)\ra \Psi(u,t)+\log q,\quad t\ra qt.
$$
This means that we can absorb $k$ into the defintion of the RG time $t$.
Therefore in the remainder of this section we set $k=1$.

\begin{figure}[tb]
\centerline{\includegraphics{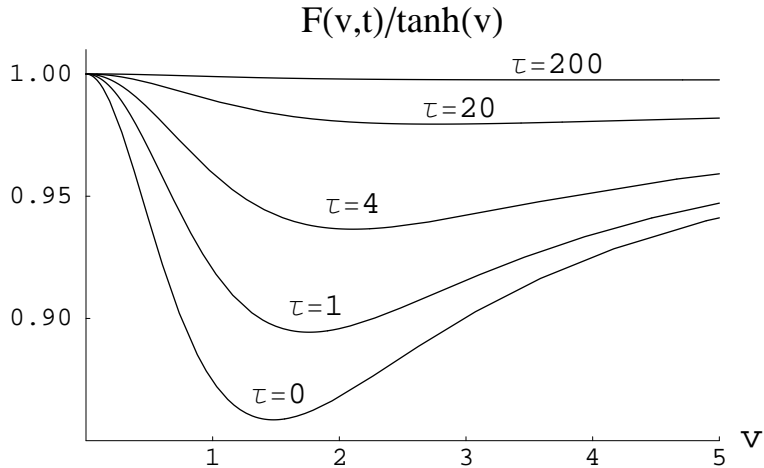}}
\caption{RG evolution of the cigar metric. We plotted $F(v,t)/\tanh v$ as
a function of $v$ for several values of the rescaled RG time $\tau=t/(4\pi)$.}
\end{figure}

For numerical integration we used an implicit scheme,
which requires solving a sparse (tri-diagonal) system of linear
equations at each step~(see e.g. \cite{Koonin}). It is also
convenient to reparametrize the variable $u$ so that it runs
over a finite rather than an infinite interval.

The results of the numerical integration of the RG equation are
presented in Figure 1. We chose to
plot the ratio $F(v,t)/\tanh v$ where $F(v,t)$ is related to 
$\Psi(u,t)$ by (\ref{FPsi}). For the 2d Black Hole this ratio
is equal to $1$. From Figure 1 it is evident that 
$F(v,t)/\tanh v$ approaches $1$ as $t\ra +\infty$. 
Hence at one-loop level
the sigma-model with target-space metric~(\ref{ours}) flows
to the 2d Black Hole~(\ref{witten}).

The discussion in this section clarifies how a linear dilaton
is generated by the RG flow. The point is simply that
as the RG time increases, the cigar tries to shrink, so that its
tip moves towards positive $u$. In order to ``keep up'' with the tip,
one has to make a $t$-dependent reparametrization of the $u$-coordinate,
which is equivalent to redefining the Weyl transformation law for
$u$. 

\section{An Exact Computation of the Central Charge}\label{ccharge}

In the previous section we have analyzed the renormalization group flow
in the one-loop approximation which is valid for large $k$.
In this section, using the method of \cite{minimal,SW},
we show that the central charge of
the IR superconformal fixed point has to be exactly
$c=3+6/k$. This computation is used in the next section to prove that
the GLSM~(\ref{action}) flows to the fermionic 2d Black Hole for all $k>0$.

\subsection{$\oQ_+$-cohomology}

One of the distinguishing properties of $(2,2)$ and $(0,2)$ theories is
the existence of topological sectors that are protected
from renormalization.
The topological sector relevant in the present context is the chiral ring,
or the right-moving chiral algebra to be more precise.
Let us choose one of the four supersymmetry generators, say $\oQ_+$.
It is a nilpotent operator whose anti-commutator with its conjugate $Q_+$
is the left-moving translation operator:
\beq
(\oQ_+)^2=0,~~~
\{\oQ_+,Q_+\}=H+P.
\eeq
By the nilpotency, one can consider $\oQ_+$ cohomology of operators.
By the second property, the left translation operator acts trivially
on the cohomology group; if $[\oQ_+,{\cal O}]=0$ then
$[H+P,{\cal O}]=\{\oQ_+,[Q_+,{\cal O}]\}\simeq 0$.
Thus correlation functions of $\oQ_+$-closed operators
are independent of $x^+=x^0+x^1$, that is, they depend only on the
$x^-=x^0-x^1$ coordinates of the insertion points.
(In the Euclidean theory they are holomorphic functions.) 
In particular they form a right-moving operator product algebra
(i.e. a chiral algebra).

Suppose a $(2,2)$ field theory flows to
a $(2,2)$ superconformal field theory.
Then $(2,2)$ supersymmetry is enhanced in
the IR limit to left-moving and right-moving
${\cal N}=2$ super-Virasoro algebras
whose generators (anti-)commute with each other.
In particular, the right-moving super-Virasoro is contained
in the chiral algebra of $\oQ_+$-cohomology classes.
By the standard argument, this ${\cal N}=2$ superconformal algebra
should be observable even at finite energy
(except in the rare case where
the IR SCFT has another copy of currents
with the same right-moving quantum numbers
but with the left-moving R-charge equal to $\pm 1$,
in which case the super-Virasoro currents
can pair up with them and disappear from the
$\oQ_+$-cohomology at finite energy).
Therefore, if one can uniquely identify such a chiral algebra
at finite energy, one can learn about the right-moving
superconformal algebra in the IR limit, and in particular compute
its central charge.

So let us look for such a superconformal algebra in the $\oQ_+$ cohomology
of the gauge theory in question.
A right-moving ${\cal N}=2$ superconformal algebra
consists of four currents
that constitute a $(2,0)$ superfield. Its
lowest component is the right-moving R-current.
What we will look for is
a $(2,2)$ superfield $\cJ$ that obeys
\beq
\bD_+\cJ=0.
\eeq
Then the lowest term in the $\theta^+, \btheta^+$ expansion
of $\cJ$ obeys the right-chiral condition
\beq
\{\oQ_+,\cJ|_{\theta^+=\btheta^+=0}\}=0,~~~
\eeq
because $\bD_+=\oQ_++2i\theta^+\partial_+$.
Hence it is a $(2,0)$ superfield that represents a $\oQ_+$ cohomology
class.
Its lowest component will flow to the right-moving R-current
of the IR theory (modulo $\oQ_+$-exact terms).
Thus, if we can identify the right-moving R-symmetry
in the high energy theory, we have a candidate
for $\cJ$.

\subsection{The current and its anomaly}

The classical system has both vector and axial $U(1)$ R-symmetries,
under which the superfields $\Phi,P$ and $\Sigma$ have charges
$(q_V,q_A)=(0,0),(0,0)$ and $(0,2),$ respectively.\footnote{
As usual, there is a room to modify the R-currents
by other global symmetries of the system. We will discuss this ambiguity in
Section \ref{subsec:ambiguity}.}
The corresponding currents are
\beqa
&&j_V^{\pm}=\opsi_{\mp}\psi_{\mp}+{k\over 2}\bchi_{\mp}\chi_{\mp}
-{1\over 2e^2}\blambda_{\mp}\lambda_{\mp},\\
&&j_A^{\pm}=\pm\psi_{\mp}\opsi_{\mp}
\pm {k\over 2}\chi_{\mp}\bchi_{\mp}
\pm{1\over 2e^2}\lambda_{\mp}\blambda_{\mp}
+{i\over e^2}(\partial_{\mp}\bsigma\sigma-\bsigma\partial_{\mp}\sigma).
\label{axial}
\eeqa
The right-moving R-current $j_R^{\pm}={1\over 2}(j_A^{\pm}-j_V^{\pm})$
is therefore expressed as
\beqa
&&j_R^+=\psi_-\opsi_-+{k\over 2}\chi_-\bchi_-
+{i\over 2e^2}\left(\partial_-\bsigma\sigma-\bsigma\partial_-\sigma\right),
\label{jR}
\\
&&j_R^-={1\over 2e^2}\blambda_+\lambda_+
+{i\over 2e^2}\left(\partial_+\bsigma\sigma-\bsigma\partial_+\sigma\right).
\eeqa
In the limit $e^2\to\infty$ where the $\Sigma$ multiplet
becomes very massive, $j_R^-$ vanishes and $j_R^+$
obeys the right-moving condition $\partial_+j_R^+=0$ classically.
Let us consider a superfield
\beqa
\cJ^{\circ}&=& D_-(\Phi\e^V)\e^{-V}\bD_-(\e^V\bPhi)
+{k\over 2}D_-(P+\bP+V)\bD_-(P+\bP+V)
\nonumber\\
&+& {i\over 2e^2}\Sigma(\partial_0-\partial_1)\bSigma.~~~~~~~~~
\label{Jcirc}
\eeqa
It is invariant under the gauge transformation (\ref{gauge}), and
its lowest component
$\psi_-\opsi_-+{k\over 2}\chi_-\bchi_-+{i\over e^2}\sigma\partial\bsigma$
is equal to $j_R^+$ up to $1/e^2$ terms.
Using the equations of motion 
\beqa
&&\bD_+\bD_-(\e^V\bPhi)=0,\label{eom1}\\
&&\bD_+\bD_-(P+\bP+V)=0,\label{eom2}\\
&&
\Phi\e^V\bPhi+{k\over 2}(P+\bP+V)
+{1\over 2e^2}(D_+\bD_-\Sigma+D_-\bD_+\bSigma)=0,
\label{eom3}
\eeqa
it is easy to check that this superfield obeys the right-chiral condition
$\bD_+\cJ^{\circ}=0$ on the classical level.

However on the quantum level this condition is violated:
\beq
\bD_+\cJ^{\circ}={1\over 2}\bD_-\Sigma.
\label{anomaly}
\eeq
This is a supersymmetric extension of the chiral anomaly
equation
\beq
\partial_{\mu}j_A^{\mu}=2F_{+-}.
\label{Aanomaly}
\eeq
The factor $2$ in front of $F_{+-}$
can be understood by noting that
there are $n$ zero modes for
{\it both} $\psi_-$ and $\opsi_+$ for a generic gauge field with
first Chern class $n=-{1\over 2\pi}\int F$.
The equation (\ref{anomaly}) is a $(1+1)$-dimensional version
of the Konishi anomaly \cite{Konishi},
and its detailed derivation is given in Appendix C.

Usually, the anomalous current cannot be modified
in a gauge-invariant way so that it is conserved.
The situation is different in the present theory
where we have a field $\vp:={\rm Im}\,p$
that shifts under the gauge transformation.
Then, the curvature $F_{+-}$ can be expressed as a
differential of a gauge invariant quantity
\beq
A_{\mu}=\partial_{\mu}\vp+v_{\mu},
\eeq
namely $F_{+-}=\partial_+v_--\partial_-v_+
=\partial_+A_--\partial_-A_+$.
Then the modified axial current
$$
\widetilde{j}_A^+=j_A^+-2A_-,\quad \widetilde{j}_A^-=j_A^-+2A_+,
$$
is gauge-invariant and conserved.
This story has a supersymmetric generalization.
Letting
\beq
\delta\cJ={1\over 2}(\bD_-D_--D_-\bD_-)(P+\bP+V),
\eeq
we can derive from (\ref{eom2})
that $\bD_+\delta\cJ=-(1/2)\bD_-\Sigma$.
This is correct quantum mechanically,
since the equation of motion (\ref{eom2}) is used linearly.
Thus the modified current
\beq
\cJ:=\cJ^{\circ}+\delta\cJ
\eeq
satisfies the right-chiral condition on the quantum level:
\beq
\bD_+\cJ=0.
\label{chirality}
\eeq
For instance, let us look at the lowest component
\beq
\cJ|_{\theta^{\pm}=\btheta^{\pm}=0}
=\psi_-\opsi_-+{k\over 2}\chi_-\bchi_-
+{i\over e^2}\sigma\partial_-\bsigma-2A_-.
\label{jm}
\eeq
{}From the chiral anomaly (\ref{Aanomaly}) and the conservation law
$\partial_{\mu}j_V^{\mu}=0$, 
it follows that $\partial_{\mu}j_R^{\mu}=F_{+-},$ or equivalently
\beq
\partial_+\left(\psi_-\opsi_-+{k\over 2}\chi_-\bchi_-
+{i\over e^2}\sigma\partial_-\bsigma\right)
+\partial_-\left({1\over 2e^2}\blambda_+\lambda_+
-{i\over e^2}\bsigma\partial_+\sigma\right)=F_{+-}=2\partial_+A_-,
\eeq
where we have used the $\vp$
equation of motion $\partial^{\mu}A_{\mu}=0$
in the last step.
We note that $\blambda_+\lambda_+-2i\bsigma\partial_+\sigma
=\{\oQ_+,\bsigma\blambda_+\}$.
Thus we find
\beq
\partial_+\left(\psi_-\opsi_-+{k\over 2}\chi_-\bchi_-
+{i\over e^2}\sigma\partial_-\bsigma-2A_-\right)=0~~
\mbox{modulo $\{\oQ_+,\cdots\}$},
\eeq
as expected from (\ref{chirality}).

\subsection{The superconformal algebra}

We define the currents $j_-,G_-,\overline{G}_-,T_-$
as the lowest components of the right-chiral
superfields $\cJ,D_-\cJ,\bD_-\cJ,{1\over 4}[\bD_-,D_-]\cJ$.
They have the following expressions in terms of component fields:
\beqa
&&j_-=\psi_-\opsi_-+{k\over 2}\chi_-\bchi_-
+{i\over e^2}\sigma\partial_-\bsigma+i(\cD_-p-\cD_-\bp),\\
&&G_-=-2i\psi_-\cD_-\bphi-ki\chi_-\cD_-\bp
+{1\over e^2}\sigma\partial_-\blambda_-+i\partial_-\chi_-,\\
&&\overline{G}_-=2i\cD_-\phi\opsi_-+ki\cD_-p\bchi_-
-{1\over e^2}\lambda_-\partial_-\bsigma-i\partial_-\bchi_-,\\
&&T_-
=2\cD_-\phi\cD_-\bphi+k\cD_-p\cD_-\bp
+{1\over 2e^2}(\partial_-\sigma\partial_-\bsigma-\sigma\partial_-^2\bsigma)
\nonumber\\[-0.15cm]
&&~~~~~~
+{i\over 2}(\psi_-\cD_-\opsi_--\cD_-\psi_-\opsi_-)
+{ik\over 4}(\chi_-\partial_-\bchi_--\partial_-\chi_-\bchi_-)
+{i\over 2e^2}\lambda_-\partial_-\blambda_-
\nonumber\\
&&~~~~~~-{1\over 2}\partial_-(\cD_-p+\cD_-\bp).
\eeqa
($j_-$ is of course identical to (\ref{jm}), as
$-2A_-=i(\cD_-p-\cD_-\bp)$.)
The quadratic terms in the currents come from $\cJ^{\circ},$ and the linear 
terms are from the ``quantum correction'' $\delta\cJ$.
Since they are the lowest components of right-chiral superfields,
they represent right-moving $\oQ_+$-cohomology classes.

Now let us compute the OPE of these currents.
We start with $j_-(x)j_-(0)$:
\beqa
&&j_-(x)j_-(0)\nonumber\\
&&\sim\psi_-\opsi_-(x)\psi_-\opsi_-(0)
+{k^2\over 4}\chi_-\bchi_-(x)\chi_-\bchi_-(0)
-{1\over e^4}\sigma\partial_-\bsigma(x)\sigma\partial_-\bsigma(0)
+4 A_-(x)A_-(0)
\nonumber\\
&&\sim
{(-i)^2\over (x^-)^2}+{k^2\over 4}{(-2i/k)^2\over (x^-)^2}
-{1\over e^4}{e^2(-e^2)\over (x^-)^2} 
+4{-1/2k\over (x^-)^2}=-{1+2/k\over (x^-)^2}.
\eeqa
Similarly, we can show that the rest of the OPE has the form
\beqa
&&j_-(x)G_-(0)\sim {-i\over x^-}G_-(0),~~
j_-(x)\overline{G}_-(0)\sim {i\over x^-}\overline{G}_-(0),
\nonumber\\
&&T_-(x)j_-(0)\sim {-1\over (x^-)^2}j_-(0)+{-1\over x^-}\partial_-j_-(0),
\nonumber\\
&&T_-(x)G_-(0)\sim {-3/2\over (x^-)^2}G_-(0)+{-1\over x^-}\partial_-G_-(0),
\nonumber\\
&&T_-(x)\overline{G}_-(0)
\sim {-3/2\over (x^-)^2}\overline{G}_-(0)
+{-1\over x^-}\partial_-\overline{G}_-(0),
\nonumber\\
&&T_-(x)T_-(0)\sim {3(1+2/k)\over 2(x^-)^4}
+{-2\over (x^-)^2}T_-(0)+{-1\over x^-}\partial_-T_-(0),
\nonumber\\
&&G_-(x)\overline{G}_-(0)
\sim {2i(1+2/k)\over (x^-)^3}
-{2\over (x^-)^2}j_-(0)
+{-2i\over x^-}\left(T_-(0)-{i\over 2}\partial_-j_-(0)\right)
\eeqa
This is an ${\cal N}=2$ superconformal algebra with central charge
\beq
c=3\left(1+{2\over k}\right).
\eeq

\subsection{Ambiguity and its resolution}\label{subsec:ambiguity}

In general, the R-current is not unique: it can be modified by other global 
symmetry currents.
This leaves an ambiguity in the definition of the R-current and therefore
in the value of the central charge.
In the present system, there is one other continuous global symmetry, namely
the shift of the imaginary part of $p$:
\beq
p\to p+i\alpha_2.
\label{shift}
\eeq
The phase rotation of $\Phi$ is another symmetry,
but that is gauge equivalent to (\ref{shift}).

The right-chiral current associated with (\ref{shift}) is given by
\beq
\cJ_2=D_-\bD_-(P+\bP+V),
\eeq
which indeed obeys $\bD_+\cJ_2=0$ by virtue of the equations of motion
(\ref{eom2}).\footnote{The chiral current for the phase rotation 
of $\Phi$ is $D_-\bD_-(\Phi\e^V\bPhi).$ This is equal to
$-{k\over 2}D_-\bD_-(P+\bP+V)+\bD_+(i\partial_-D_-\bSigma/e^2)$
by the equation of motion
(\ref{eom3}). Therefore this current is proportional to $\cJ_2$
modulo $\bD_+$ exact terms.}
This current is free of Konishi anomaly or
$\oQ_+$-anomaly in the sense of \cite{SW},
because the conservation equation $\bD_+\cJ_2=0$ is derived by using the 
equation of motion linearly.
Thus it appears that one can modify the current $\cJ$ by an arbitrary 
multiple of $\cJ_2$
\beq
\cJ'=\cJ+a\cJ_2.
\eeq
It is easy to see that the four currents $j'_-,G'_-,\overline{G}'_-,T'_-$
defined as above form an ${\cal N}=2$ superconformal algebra
with a central charge $c'=3+6(1-a)/k$.
Which of these $\cJ'$ yields the 
superconformal algebra in the infrared limit?
Since the central charge has to be real, we know that $a$ is real,
but we still have an ambiguity.

One can fix this ambiguity using a mild assumption
about the low energy limit of the theory.
Let us look at the expression for $j'_-$:
\beqa
j'_-&=&\psi_-\opsi_-+{k\over 2}\chi_-\bchi_-
+{i\over e^2}\sigma\partial_-\bsigma+i(\cD_-p-\cD_-\bp)+ia\cD_-\bp,
\nonumber\\
&=&{\rm Re}\,j'_-+i\left({1\over 2e^2}\partial_-|\sigma|^2+a\partial_-\rp
\right).
\label{imaginary}
\eeqa
The current in the infrared limit
has to be real and therefore
the imaginary part in (\ref{imaginary}) has to vanish
up to $\oQ_+$-exact terms. The mild assumption is the existence of the 
asymptotic region
at $\rp\to -\infty$ where the theory flows to the sigma-model on a flat
cylinder, possibly with a linear dilaton of {\it some} slope.
The term ${1\over 2e^2}\partial_-|\sigma|^2$
is negligible in that region, because $\sigma$ has a large mass due
to large values of $|\phi|^2\sim -\rp/2$.
On the other hand, the field $\partial_-\rp$ survives in the IR limit as a 
free field (possibly with a background charge), and is not $\oQ_+$-exact.
Thus for the current to be real up to $\oQ_+$-exact terms, we have to set
\beq
a=0.
\eeq
It follows that $j_-,G_-,\overline{G}_-,T_-$ are the unique currents
with the right properties, and the central
charge of the IR fixed point is exactly $c=3+6/k$.
Note that the slope of the linear dilaton
is uniquely fixed by the chiral anomaly.

\section{Flow to 2d Black Hole II: Exact Treatment}\label{flowtwo}

In the previous sections, we have seen that
the gauged linear sigma model ~(\ref{action}) flows to a $(2,2)$ 
superconformal field theory with the same central charge, symmetries,
and asymptotic behavior as the fermionic 2d Black Hole. 
However, there remains a possibility that
it flows not to the supercoset itself, but
to some other nearby fixed point with the same properties.
The goal of this section is to argue that this
does not happen.

\subsection{General remarks}

The fermionic 2d Black Hole is defined as the
supersymmetric $SL(2,\RR)/U(1)$ coset at level $k$. The 
central charge is 
\begin{equation}\label{centralcharge}
c=3\left(1+\frac{2}{k}\right).
\end{equation}
Unlike in the bosonic case, here the expansion of the central charge
in powers of $1/k$ terminates at one-loop order. For large
$k$ this CFT is weakly coupled and is equivalent to the sigma-model 
with target~(\ref{witten}). Note that the central charge of the
fermionic 2d Black Hole at level $k$ is exactly the same as
the IR central charge of the GLSM~(\ref{action})
computed in Section~\ref{ccharge}. In the asymptotic region of the target
space both models become equivalent to the theory of a free chiral
superfield with radius $\sqrt {k}$ and a background charge. 
The $SL(2,\RR)/U(1)$ supercoset is
an example of a Kazama-Suzuki model and has $(2,2)$ supersymmetry.
The world-sheet parity is also a symmetry of the model (see Appendix D).
There is also a global non-R symmetry, the 
momentum symmetry (this is clear from the fact that the sigma-model
metric~(\ref{intro:cigar}) describing the supercoset has a
$U(1)$ isometry which shifts $\varphi$).
Thus the supercoset has the same symmetries as the IR fixed point of the
GLSM.

The analysis of Section~\ref{flowone} shows that for $k\ra\infty$ 
the GLSM~(\ref{action}) flows to the fermionic 2d Black Hole at level $k$. 
For finite $k$ we only know that the GLSM flows to a $(2,2)$ 
superconformal field theory with the same central charge, symmetries,
and asymptotic behavior as the fermionic 2d Black Hole at level $k$. 
It could be that for finite $k$ the GLSM flows not to the supercoset, but 
to some other fixed point nearby. But if this is the case, then the supercoset 
theory must admit a marginal operator which deforms it to the IR fixed point
to which the GLSM flows to. This operator must preserve all the symmetries
of the 2d Black Hole and leave its asymptotic behavior unchanged.
If we can show that such marginal operators are absent, then the
GLSM~(\ref{action}) has no choice but to flow to the fermionic 2d
Black Hole for all $k>0$.

\subsection{Marginal deformations of the bosonic coset}

As a warm-up, let us discuss marginal deformations of the bosonic 
$SL(2,\RR)/U(1)$ coset. This problem has been previously addressed
in~\cite{DVV,DN}. We will focus on marginal deformations which preserve
all the obvious symmetries of the coset, i.e. momentum and world-sheet
parity. In addition we require the deformation to decay or stay constant
towards $\rho\ra \infty$, so that the asymptotic behavior of the model
is not drastically altered.

First, let us consider marginal operators in the coset which correspond to 
normalizable states in the parent WZW theory.
The quantization of the $SL(2,\RR)$ WZW has been a subject of interest
for many years, but the precise spectrum of the theory was determined
only recently~\cite{MO}. According to~\cite{MO}, one should include the
following representations of $SL(2,\RR)$ as the Kac-Moody primaries:

(i) $\cD^+_j$: principal discrete representation with lowest weight
of spin $j$, ${1\over 2}<j<{k-1\over 2}$.

(ii) $\cD^-_j$: principal discrete representation with highest weight
of spin $-j$, ${1\over 2}<j<{k-1\over 2}$.

(iii) ${\cal C}^\alpha_j$: principal continuous representations with 
$j=\frac{1}{2}+is,\ s\in\RR$ and $0\leq \alpha<1.$

\noindent
We will work on the universal cover of $SL(2,\RR)$, in which case $\rre\ j$
is not quantized. 
The primaries transforming in the principal discrete representations are 
normalizable, while the primaries in the principal continuous 
representations are delta-function normalizable. 

As usual, positive-energy representations of the $SL(2,\RR)$ current 
algebra are obtained by declaring that $J^\pm_n, J^3_n$ annihilate the 
primaries for all $n>0$. We denote these representations 
by ${\wh\cD}^\pm_j,{\wh{\cal C}}^\alpha_j.$ However,
one should also include other representations labeled by an
integer $w$~\cite{MO}.
These are obtained by declaring that the primary is
annihilated by $J^+_{n+w},J^3_n,$ and $J^-_{n-w}$ for $n>0$. 
One says that these new representations are obtained from the usual
positive-energy representations by the spectral flow. They are denoted
by ${\wh\cD}^{\pm,w}_j$ and ${\wh{\cal C}}^{\alpha,w}_j$.
Under the spectral flow by $w$ units,
the $L_0$ and $J_0^3$ eigenvalues of a state change as
$(h,m)\mapsto (h+wm-kw^2/4,m-kw/2)$.
In general spectral flow takes a positive-energy representation of
$\wh{SL(2,\RR)}$ to a representation with energy unbounded from below.
The exceptions to this rule are $\wh\cD^{+,w=-1}_j$ and
$\wh\cD^{-,w=1}_j$. They are equivalent to $\wh\cD^-_{\frac{k}{2}-j}$
and $\wh\cD^+_{\frac{k}{2}-j}$, respectively. More generally, we have
\beq\label{isomorphism}
\wh\cD^{-,w}_j\simeq \wh\cD^{+,w-1}_{\frac{k}{2}-j}.
\eeq
Hence, to avoid double-counting, we should include in the spectrum 
$\wh\cD^{+,w}_j$ and $\wh{\cal C}^{\alpha,w}$ for all $w\in\ZZ$, but 
exclude $\wh\cD^{-,w}_j$.

The amount of spectral flow in the left-moving
and right-moving sectors must be the same~\cite{MO}.
Thus the space of states of the $SL(2,\RR)$ WZW model at level $k$ is
the sum of $\wh{\cD}^{+,w}_j\times \wh{\cD}^{+,w}_j$
($1<j<{k-1\over 2}$, $w\in\ZZ$) and
$\wh{\cal C}^{\alpha,w}_j\times \wh{\cal C}^{\alpha,w}_j$
($j\in {1\over 2}+i\RR$, $w\in\ZZ$).
Before the spectral flow the spin-$j$ primary state
with $J^3_0=m,\tJ^3_0=\tm$ has conformal weights
\beq
L_0=\tL_0=-\frac{j(j-1)}{k-2}.
\eeq
After the spectral flow by $w$ its quantum numbers become
\beqa
&&J_0^3=m-{kw\over 2},~~\tJ_0^3=\tm-{kw\over 2},
\label{wJ0}\\
&&L_0=-\frac{j(j-1)}{k-2}+wm-\frac{kw^2}{4},\quad
\tL_0=-\frac{j(j-1)}{k-2}+w\tm-\frac{kw^2}{4}.
\label{wL0}
\eeqa

States of the coset theory are represented by states of the parent WZW
theory obeying
\beqa
&&J_0^3+\tJ_0^3=0,\\
&&J_n^3=\tJ_n^3=0,~~n\geq 1.
\label{jnp}
\eeqa
The momentum in the coset theory is given by
$$
J_0^3-\tJ_0^3.
$$
The Virasoro generators are represented by
$L_n-L_n^{U(1)}$,
$\tilde{L}_n-\tilde{L}_n^{U(1)}$ where
$L_n^{U(1)}$ and $\tilde{L}_n^{U(1)}$ are the Sugawara operators of the
$U(1)$ subalgebra at level $k$.

We are interested in Virasoro primaries in the coset theory which have
dimension $(1,1)$ and zero momentum. This means that we are looking for 
Virasoro primaries of the parent WZW theory satisfying (\ref{jnp}) 
together with 
\beq
J_0^3=\tilde{J}_0^3=0~~\mbox{and}~~L_0=\tilde{L}_0=1.
\eeq
A little high-school algebra shows that in the discrete representations
there are two such states for $k>3$:
\beqa\label{firststate}
&& J^-_{-1}\tJ^-_{-1}|j=1\rangle^{+},
\\
&& \left[J_0^+\tJ_0^+\vert j=\mbox{${k\over 2}-1$}\rangle^+\right]^{w=-1}.\label{secstatep}
\eeqa
Here $|j\rangle^{\pm}$ is the lowest/highest weight primary state
of $\wh{\cD}^{\pm}_j\times\wh{\cD}^{\pm}_j$,
and $[-]^w$ is the spectral flow of $[-]$ by $w$ units. These two states
are related by world-sheet parity. This becomes clearer if we use the
isomorphism of ${\wh\cD}^{+,w=-1}_{\frac{k}{2}-1}$ and ${\wh\cD}^-_1$
and write the second state as\footnote{This should not be confused
with the ``field identification'' in coset models \cite{FI}
which would happen only if the gauge group had a non-trivial
fundamental group \cite{FIH}.
We are considering the universal cover of
$SL(2,\RR)$ modded out by the gauge group $\RR$.
Since $\pi_1(\RR)=\{1\}$,
there is no non-trivial field identification.}

\beq\label{secondstate}
J^+_{-1}\tJ^+_{-1}|j=1\rangle^{-}.
\eeq
Since world-sheet parity exchanges $J^{\pm}$ and $\tJ^{\mp}$ and
$\wh{\cD}_j^+\times \wh{\cD}_j^+$ and
$\wh{\cD}_j^-\times\wh{\cD}_j^-$
(see Appendix \ref{app:D}), the statement becomes
obvious.

The above two states are in the spectrum if $1<(k-1)/2$, i.e. for $k>3$.
For $k=3$ the states become delta-function normalizable and appear in the
continuous representations with $j=1/2,\alpha=1/2,w=\pm 1$ (see below). 
For $2<k<3$ the states are not normalizable.

Thus for $k>3$ there are two marginal operators in the $SL(2,\RR)$ WZW
theory which come from discrete representations and could give rise 
to marginal momentum-conserving deformations of the coset. 
It is easy to write down their explicit form.
Following \cite{MO} we use the coordinates $(\rho,t,\varphi)$ on $SL(2,\RR)$
defined by
\beq
g=\e^{i\sigma_2(t+\varphi)/2}
\e^{\sigma_3\rho}\e^{i\sigma_2(t-\varphi)/2}
\eeq
($\phi$ of \cite{MO} is replaced here with $\varphi$
to avoid confusion with the scalar component of $\Phi$).
The vertex operators corresponding to the
two states in~(\ref{firststate},\ref{secondstate}) are given by
\beq\label{bosvertex}
\left(\frac{\partial_+\rho}{\cosh\rho}\mp i\sinh\rho\
\partial_+(-t-\varphi)
\right)\cdot
\left(\frac{\partial_-\rho}{\cosh\rho}\mp 
i\sinh\rho\ \partial_-(-t+\varphi)\right).
\eeq
They are complex-conjugates of each other and are exchanged by world-sheet 
parity $\partial_+\leftrightarrow \partial_-$, $t\leftrightarrow -t$
(Appendix \ref{app:D}).

The coset can be realized as a gauged WZW model~\cite{2dbh,DVV}.
The gauging is with respect to the translation symmetry
$t\to t-\alpha$, and the gauged action is
obtained from the ordinary WZW action by 
replacing $\partial_{\mu}t$ with the gauge invariant
expression $\partial_{\mu}t-A_{\mu}$:
\beqa
S&=&kS_{\rm WZW}(A,g)
\label{gWZW}\\[0.1cm]
&=&{k\over 4\pi}\int\dd^2x\Biggl[
-\eta^{\mu\nu}\Bigl(
\partial_{\mu}\rho\partial_{\nu}\rho
+\sinh^2\!\!\rho\,\partial_{\mu}\varphi\partial_{\nu}\varphi
-\cosh^2\!\!\rho\,\partial_{\mu}t\partial_{\nu}t\Bigr)
\nonumber\\
&&~~~~~~~~~~~~~
-4\sinh^2\!\!\rho\,(\partial_-t\partial_+\varphi-\partial_+t\partial_-\varphi)
-4\cosh^2\!\!\rho \,A_+A_-
\nonumber\\
&&~~~~~~~~~~~~~
+4(\cosh^2\!\!\rho\,\partial_-t-\sinh^2\!\!\rho\,\partial_-\varphi)A_+
+4(\cosh^2\!\!\rho\,\partial_+t+\sinh^2\!\!\rho\,\partial_+\varphi)A_-
\Biggr].
\nonumber
\eeqa
The vertex operators in the coset model corresponding to
the states (\ref{firststate},\ref{secondstate}) are obtained from 
(\ref{bosvertex})
by replacing
$\partial_{\mu}t\to\partial_{\mu}t-A_{\mu}$.
The equations of motion for $A_{\mu}$ imply
\beq
A_--\partial_-t
=-\tanh^2\!\!\rho\,\partial_-\varphi,
~~
A_+-\partial_+t=\tanh^2\!\!\rho\,\partial_+\varphi.
\eeq
Substututing these expressions into (\ref{gWZW}),
we obtain the world-sheet Lagrangian for the cigar (\ref{intro:cigar}).
(The variable $t$ completely
disappears from the action and can be
ignored.)
Substuting the same expressions into
the gauged versions of (\ref{bosvertex}), 
we see that the vertex operators reduce to
\beq\label{bosotwo}
\frac{1}{\cosh^2\!\!\rho\,}\Bigl(\partial_+\rho\partial_-\rho
+\tanh^2\!\!\rho\, 
\partial_+\varphi\partial_-\varphi\Bigr)
\pm i
{\tanh\rho\over \cosh^2\!\!\rho\,}
(\partial_-\rho\partial_+\varphi-\partial_+\rho\partial_-\varphi).
\eeq
The real part is a metric deformation,
and at first sight it seems non-trivial, but
in fact it is a total derivative on the world-sheet. To see this, note
that an infinitesimal reparametrization of the $\rho$ coordinate,
$\rho'=\rho+\eps\tanh\rho$, changes the metric of the cigar by
$$
\frac{2\eps}{\cosh^2\rho}\left(\dd\rho^2+\tanh^2\rho\ \dd\varphi^2\right).
$$
Thus deformation by the real part of~(\ref{bosotwo}) is equivalent to a 
reparametrization of $\rho$. This implies in turn that this deformation
is a total derivative on the world-sheet. Using equations of motion, one
can check that~(\ref{bosotwo}) is proportional to
\beq\label{totderiv}
\partial_+\partial_- \log \cosh^2\rho.
\eeq
The imaginary part of (\ref{bosotwo}) is a B-field term which is 
parity-odd.
This, of course, corresponds to the fact that the two states in
(\ref{firststate},\ref{secondstate}) are exchanged by world-sheet parity.

The conclusion is that for $k>3$ the discrete series give rise to 
two momentum-conserving
marginal deformations in the coset theory (while for $k\leq 3$ they give none).
One is a parity-odd B-field, and the other is
a total derivative on the world-sheet. If we restrict ourselves
only to parity-even deformations, then we are left with the total derivative
operator. Can one simply discard this operator as trivial? If the world-sheet
is compact without a boundary, then one is certainly justified in doing so,
but if the world-sheet has a boundary, or is noncompact, like $\RR^2$,
then the answer depends on boundary conditions. Since
we are studying a conformal field theory, it is natural to impose boundary
conditions which preserve Weyl invariance. In this case, the total derivative
operator is trivial. Indeed, recall that the variables $u$ and $\rho$ are
related by $u=\log\sinh\rho$, and therefore a change of variables
$\rho\ra \rho+\eps\tanh\rho$ is equivalent to $u\ra u+\eps$. But the latter
change of variables is also effected by the Weyl transformation. Hence
with Weyl-invariant boundary conditions adding the operator~(\ref{totderiv})
has no effect on the theory.

As for the principal continuous series, for general $k$ the only states that 
give rise to marginal operators in the coset theory are the Kac-Moody primaries with 
\beq\label{cont}
j=\frac{1}{2}\pm i\sqrt{k-\frac{9}{4}},\quad m=\bm=0.
\eeq
Such operators decay as $\exp(-2j\rho)$, and since
for $k>9/4$ $j$ has a nonzero imaginary part, they exhibit oscillatory
behavior. Note that for 
$k=\frac{9}{4}$ there appears a non-oscillatory vertex operator decaying
as $e^{-\rho}$. This is related to the fact that for $k=9/4$ the central
charge of the coset is 26, and the ``tachyon'' in the corresponding
critical string theory is massless. The above vertex operator then
describes the emission of the zero mode of the tachyon~\cite{DVV}.

In addition, for $k=3$ there appear two additional $(1,1)$ states
in $\wh{{\cal C}}^{\alpha=\frac{1}{2},w=\pm 1}_{\frac{1}{2}}$: 
\beqa\label{contkthree}
&& \left[J^+_0 \tJ^+_0 \vert \ts j=\frac{1}{2},
\alpha=\frac{1}{2}\rangle^+\right]^{w=1},\\
&& \left[J^-_0 \tJ^-_0 \vert \ts j=\frac{1}{2},
\alpha=\frac{1}{2}\rangle^-\right]^{w=-1}.\nonumber
\eeqa
Note that for $\alpha=1/2$ the continuous representation becomes reducible
and decomposes into a direct sum of a highest weight representation with
highest spin $-1/2$ and a lowest weight representation with lowest spin $1/2$.
This explains superscripts $\pm$ in the above formula.
The states~(\ref{contkthree}) can be regarded as the $k\ra 3$ limit of the 
discrete states~(\ref{firststate},\ref{secondstate}). To see this, one should use
the isomorphism~(\ref{isomorphism}). 

Now let us turn to operators corresponding to non-normalizable
states. Recall that the zero-mode wave-functions of primary states 
with spin $j$ decay as $\exp(-2j\rho)$~\cite{MO}. 
Since the volume element of $SL(2,\RR)$ is proportional to
$$
\sinh 2\rho\ dt\ d\varphi,
$$
the zero-mode wave-function is normalizable for $\rre\ j>1/2$. 
This was the origin of
the restriction $j>1/2$ for the discrete series. If we do not require
normalizability, but would like the wave-function to decay towards
$\rho\ra \infty$ or at least not to grow, we can relax this constraint to $j\geq 0$. 
Consistency with
the spectral flow then requires
$0\leq j\leq k/2$. This is to be compared with the normalizability condition
$\frac{1}{2} < j < \frac{k-1}{2}.$ 
It is easy to check that relaxing the conditions on $j$ gives
just one extra $(1,1)$ state, namely the one with $j=0,m=\bm=0, w=0$
and vertex operator
$$
J^3(x^-)\tJ^3(x^+).
$$
But this operator becomes zero after passing to the coset theory. 

In addition, relaxing the constraint on $j$ has the following effect.
Recall that for $2<k<3$ the states with $j=1$, and in particular 
the $(1,1)$ states (\ref{firststate},\ref{secondstate}), are not in the 
spectrum.
With the relaxed constraint $0<j<k/2$ these two operators are allowed all 
the way down to $k=2$. As explained above, the parity-even combination of 
the two operators is trivial if Weyl-invariant boundary conditions are used on
the world-sheet, while the parity-odd one is a B-field.

To summarize, the only non-trivial marginal deformation of the bosonic coset 
which preserves
momentum and world-sheet parity is a tachyon potential corresponding
to the parity-even combination of the states~(\ref{cont}). This operator
exists for all $k\geq 9/4$ and is delta-function normalizable. For $k=9/4$
it becomes the usual Liouville potential deformation. This result is somewhat 
puzzling from the perspective of the FZZ duality conjecture. According to~\cite{KKK},
the sine-Liouville theory admits at least two marginal deformations: the
Liouville potential and the radius-changing operator. In the coset theory
we see the former, but no trace of the latter. However, it seems 
plausible that the deformation of the supercoset which changes the
asymptotic radius of the cigar leads to a conical singularity at $\rho=0$.
The above analysis assumes from the beginning that the deformation 
is everywhere smooth and therefore cannot detect the radius-changing
operator.\footnote{We are grateful to Steve Shenker and Juan Maldacena for
emphasizing this point to us.} Since the bosonic
FZZ duality is not the subject of this paper, we will not dwell any further
on this issue. The situation in the supersymmetric case is somewhat different,
as discussed below.

\subsection{Marginal deformations of the supersymmetric coset}

We now analyze marginal deformations
of the Kazama-Suzuki supersymmetric coset model which preserve 
$(2,2)$ supersymmetry, R-symmetry and world-sheet parity.
The model is defined as the $SL(2,\RR)$ WZW model
at level $(k+2)$ plus a Dirac fermion,
modded out by a $U(1)$ acting on the $SL(2,\RR)$ part
as before and axially on the fermion \cite{KazS}.
Thus the analysis is different from the bosonic case
by a shift of the level $k\to k+2$
and by the addition of the fermionic sector.
Fermionic oscillators
$\psi_r,\opsi_r$ (right-moving) and
$\tpsi_r,\overline{\tpsi}_r$ (left-moving)
have the following commutation relation with $J_0^3$ and $\tJ_0^3$:
\beq
\begin{array}{cc}
{[}J_0^3,\psi_r]=-\psi_r,&
{[}J_0^3,\opsi_r]=\opsi_r,\\[0.1cm]
{[}\tJ_0^3,\tpsi_r]=\tpsi_r,&
{[}\tJ_0^3,\overline{\tpsi}_r]=-\overline{\tpsi}_r.
\end{array}
\label{fermions}
\eeq
For our purpose, we can work in the NS-NS sector,
$r\in {1\over 2}+\Z$,
which has a vacuum $|0\rangle$ that is annihilated by
oscillators of positive frequency modes
(and therefore is also annihilated by $J_0^3,\tJ_0^3$).

The state space of the theory before the gauging of $U(1)$
is a tensor product of the state space of the
$SL(2,\RR)_{k+2}$ WZW model and the Fock space ${\cal F}$ of the Dirac fermion.
The former is the same as in the bosonic case with a shift of the level 
$k\to k+2$.
(The representation $\wh{\cD}^{\pm}_j$ is now isomorphic to
$\wh{\cD}^{\mp,w=\pm 1}_{{k\over 2}+1-j}$.)
The spectral flow acts on the fermions as well as the bosons and
sends the Fock space ${\cal F}$ to itself. In particular,
\beq
|0\rangle
\begin{array}{c}
-w\\[-0.35cm]
\longrightarrow\\[-0.35cm]
\longleftarrow\\[-0.35cm]
w
\end{array}
\left\{
\begin{array}{ll}
\opsi_{-w+{1\over 2}}\cdots\opsi_{-{1\over 2}}
\tilde{\psi}_{-w+{1\over 2}}\cdots\tilde{\psi}_{-{1\over 2}}|0\rangle,
&w\geq 1,\\
\psi_{-|w|+{1\over 2}}\cdots\psi_{-{1\over 2}}
\overline{\tilde{\psi}}_{-|w|+{1\over 2}}\cdots
\overline{\tilde{\psi}}_{-{1\over 2}}|0\rangle,
&w\leq -1.
\end{array}
\right.
\label{ide}
\eeq
We can regard the total state space as
the tensor sum of
\beqa
&&(\wh{\cD}^{+}_j\times\wh{\cD}^{+}_j)\otimes {\cal F}
~~~~\mbox{(${1\over 2}<j<{k+1\over 2}$)},~~\mbox{and}
\label{cDpm}\\
&&(\,\wh{\cal C}^{\alpha}_{j}\,\times\,
\wh{\cal C}^{\alpha}_{j}\,)\otimes {\cal F}
~~~~\mbox{($j\in {1\over 2}+i\RR$, $0\leq\alpha <1$)},
\label{Calpha}
\eeqa
and their spectral flows.
Before the spectral flow, the spin $j$ primary state
with $J_0^3=m,\tJ_0^3=\tm$ 
has conformal weights
\beq
L_0=\tilde{L}_0=-{j(j-1)\over k}.
\eeq
After the spectral flow by $w$ units, it becomes a state with
\beqa
&&J_0^3=m-{kw\over 2},~~\tJ_0^3=\tm-{kw\over 2},
\label{wJ}\\
&&L_0=-{j(j-1)\over k}+wm-{k\over 4}w^2,~~
\tL_0=-{j(j-1)\over k}+w\tm-{k\over 4}w^2.
\label{wL}
\eeqa
Despite the level shift $k\to k+2$, the coefficient of $w$ in (\ref{wJ})
and $w^2$ in (\ref{wL}) is proportional to $k,$ as in (\ref{wJ0})
and (\ref{wL0}), because the fermionic sector contributes $-2$.

States of the coset model must obey
$J_0^3+\tJ_0^3=0$ and
$J_n^3=\tJ_n^3=0$ for $n\geq 1$. The momentum generator is
given by $J_0^{3(b)}-\tJ_0^{3(b)}$, where $J_0^{3(b)}$ is the bosonic
part of the $SL(2,\RR)$ generator $J_0^3$.
The Virasoro generators are defined as usual.
There are also $(2,2)$ superconformal generators defined as follows 
\cite{KazS}:
\beq
\begin{array}{l}
G_r\propto\sum_{n\in \Z}\psi_{r+n}J^+_{-n},\\[0.15cm]
\overline{G}_r\propto\sum_{n\in\Z}\opsi_{r+n}J^-_{-n},\\[0.15cm]
J_n\propto J_n^{3(f)}+{2\over k+2}J_n^{3(b)},
\end{array}
~~
\begin{array}{l}
\tG_r\propto\sum_{n\in\Z}\tpsi_{r+n}\tJ^-_{-n},\\[0.15cm]
\overline{\tG}_r\propto
\sum_{n\in\Z}\overline{\tpsi}_{r+n}\tJ^+_{-n},\\[0.15cm]
\tJ_n\propto \tJ_n^{3(f)}+{2\over k+2}\tJ_n^{3(b)},
\end{array}
\label{supercu}
\eeq
where $J_n^{3(f)}$ and $J_n^{3(b)}$ are fermionic and bosonic parts
of $J_n^3=J_n^{3(f)}+J_n^{3(b)}$.

We would like to find all even states in the coset model
with zero axial and vector charges which preserve momentum and 
whose integral over
the world-sheet is a $(2,2)$ superconformal invariant. The last requirement 
means that
they must be Virasoro primaries of weight $(1,1)$ and be annihilated by 
$G_{-1/2},\tG_{-1/2}$ up to total derivatives.
Momentum conservation requires
$J_0^{3(b)}-\tJ_0^{3(b)}=0$, and together with
R-invariance this implies that the fermionic and bosonic parts of
$J_0^3,\tJ_0^3$ have to vanish independently.

We start with the discrete representations and their spectral flows.
A little more high-school algebra reveals that the only normalizable
states satisfying the above requirements are
\beqa
&&J^-_{-1}\tilde{J}^-_{-1}|j=1\rangle^+\otimes |0\rangle,
\label{st1}\\
&&J^+_{-1}\tilde{J}^+_{-1}|j=1\rangle^-\otimes |0\rangle.
\label{st2}
\eeqa
They can be equivalently written as
\beqa
&&\left[J_0^+\tilde{J}_0^+|j=\mbox{${k\over 2}$}\rangle^+\otimes
\psi_{-{1\over 2}}\overline{\tilde{\psi}}_{-{1\over 2}}
|0\rangle\right]^{w=-1},\\
&&\left[J_0^-\tilde{J}_0^-|j=\mbox{${k\over 2}$}\rangle^-\otimes
\opsi_{-{1\over 2}}\tilde{\psi}_{-{1\over 2}}
|0\rangle\right]^{w=1}.
\eeqa
These states are in the spectrum for $k>1$.
They are supersymmetry-descendants of $({1\over 2},{1\over 2})$ primary 
states.
Namely (\ref{st1}) and
(\ref{st2})
can be expressed respectively as
$\overline{G}_{-{1\over 2}}\tG_{-{1\over 2}}$ and
$G_{-{1\over 2}}\overline{\tG}_{-{1\over 2}}$ applied to
the $({1\over 2},{1\over 2})$ states
\beq
|j=1\rangle^+\otimes 
\psi_{-{1\over 2}}\overline{\tpsi}_{-{1\over 2}}|0\rangle,
~~\mbox{and}~~
|j=1\rangle^-\otimes \opsi_{-{1\over 2}}\tpsi_{-{1\over 2}}|0\rangle.
\eeq
Furthermore, these two $({1\over 2},{1\over 2})$ states are primaries of the
$(2,2)$ superconformal algebra
annihilated by $G_{-{1\over 2}},\overline{\tG}_{-{1\over 2}}$ and
$\overline{G}_{-{1\over 2}},\tG_{-{1\over 2}},$ respectively,
and therefore are twisted (anti-)chiral
primaries.
Thus, the integrals of operators corresponding to
(\ref{st1}) and (\ref{st2})
are twisted F-terms. Since they have vanishing R-charges, they are in
fact exactly marginal deformations of the supercoset theory.
However, as in bosonic case, the parity-even combination of these
operators is essentially trivial. We now explain this.

A Kazama-Suzuki supercoset can be realized as a supersymmetric gauged WZW
model~\cite{Nmatrix}.
The Dirac fermion transforms under the $U(1)$ gauge group as
$\psi_{\mp}\to \e^{\mp i\alpha}\psi_{\mp}$ and
$\opsi_{\mp}\to\e^{\pm i\alpha}\opsi_{\mp}$. This is equivalent 
to~(\ref{fermions}), if $\psi_r$ and $\tpsi_r$ are the
modes of $\psi_-$ and $\psi_+,$ respectively.
The action is given by
\beqa
S&=&(k+2)S_{\rm WZW}(A,g)
\nonumber\\
&&+{1\over 2\pi}\int \dd^2x\left[
2i\opsi_-(\partial_+-iA_+)\psi_-
+2i\opsi_+(\partial_-+iA_-)\psi_+
\right].
\label{saction}
\eeqa
The states (\ref{st1}) and (\ref{st2}) are identical to 
(\ref{firststate},\ref{secondstate}) up to tensor product with the 
vacuum vector of the fermionic Fock space.
Thus, the vertex operators for the former states are still given
by (\ref{bosvertex}). The difference between the bosonic and supersymmetric
cases arises only after gauging.
The equation of motion for $A_{\mu}$ is solved by
\beq
A_{\mp}-\partial_{\mp}t
=\mp \tanh^2\!\!\rho\,\partial_{\mp}\varphi
\pm {1\over (k+2)\cosh^2\!\!\rho\,}\opsi_{\mp}\psi_{\mp}.
\label{Aexp}
\eeq
Substituting this into the action (\ref{saction}),
we obtain the Lagrangian
\beqa
&&-{k+2\over 2}\eta^{\mu\nu}\Bigl(
\partial_{\mu}\rho\partial_{\nu}\rho
+\tanh^2\!\!\rho\,\partial_{\mu}\varphi\partial_{\nu}\varphi\Bigr)
\nonumber\\
&&
+2i\opsi_-(\partial_+-i\partial_+t
-i\tanh^2\!\!\rho\,\partial_+\varphi)\psi_-
+2i\opsi_+(\partial_-+i\partial_-t
-i\tanh^2\!\!\rho\,\partial_-\varphi)\psi_+
\nonumber\\
&&
-{2\over (k+2)\cosh^2\!\!\rho\,}
\psi_+\psi_-\opsi_-\opsi_+,
\eeqa
which describes the supersymmetric cigar.
\footnote{It can also be written as
${k+2\over 2}\int \dd^4\theta\,K(Z,\bar Z)$
where $Z$ is a chiral superfield with components
$$
z=\log\sinh\rho+i\varphi,~~
\chi_{\pm}=\sqrt{2\over k+2}\coth\rho\e^{-i\varphi\pm it}\psi_{\pm},
$$
and the K\"ahler potential is such that $K_{z\bar z}=1/(1+|\e^{-z}|^2)$.
This shows that (\ref{supercu}) is in the standard convention
with respect to chiral versus twisted chiral.
$z$ is a good variable away from the tip of the cigar.
A good coordinate near the tip is $w=\e^z$, with
$K_{w\bar w}=1/(1+|w|^2)$.}
Substituting (\ref{Aexp}) into the gauged version of
(\ref{bosvertex}), we obtain explicit expressions for vertex operators
in the supercoset:
\beq\label{supervertexop}
\frac{1}{\cosh^2\!\!\rho\,}\Bigl(\partial_+\rho\partial_-\rho
+\tanh^2\!\!\rho\, 
\widetilde{\partial_+\varphi}\widetilde{\partial_-\varphi}\Bigr)
\pm i
{\tanh\rho\over \cosh^2\!\!\rho\,}
(\partial_-\rho\widetilde{\partial_+\varphi}
-\partial_+\rho\widetilde{\partial_-\varphi}),
\eeq
where we denoted 
\beq
\widetilde{\partial_{\mp}\varphi}
=\partial_{\mp}\varphi+{1\over k+2}\opsi_{\mp}\psi_{\mp}.
\eeq
The real part is proportional to the variation of the action under the
change of variables $\delta\rho=\eps\tanh\rho$. As in the bosonic case, this
means that this deformation is trivial if the world-sheet is compact,
or if Weyl-invariant boundary conditions are imposed on the world-sheet
boundary.
The imaginary part corresponds to switching on the B-field.
It is parity-odd, in agreement with the fact that
the two states (\ref{st1}) and (\ref{st2}) are exchanged by world-sheet
parity. 

One can in fact identify both of the above deformations 
in the gauged linear sigma-model: they are 
the Fayet-Iliopoulos term and the theta-angle:
\beq\label{FItheta}
{\rm Re}\int\dd^2\widetilde{\theta}\,
(r-i\theta)\Sigma.
\eeq
The Fayet-Iliopoulos deformation is trivial as it can be absorbed into
the real part of $P$, while the theta-angle breaks world-sheet parity. 
One can easily check that in the presence of the theta-angle integrating out 
the gauge field yields a supersymmetric sigma-model with target 
metric~(\ref{ourcoset}) and a B-field
\beq\label{bfield}
B_{\rho\varphi}\sim \frac{\tanh\rho}{\cosh^2\rho}.
\eeq
Topological terms in the action, like the B-field with vanishing $H=dB$, 
are not subject to RG flow. Therefore it is gratifying that the expression 
(\ref{bfield}) obtained by a classical computation agrees with the
imaginary part of (\ref{supervertexop}).

Next we consider continuous representations and their spectral flow.
It is easy to see that the only states obeying
$J_0^{3(b,f)}=\tJ_0^{3(b,f)}=0$ and $L_0=\tL_0=1$ are
the primary states $|j\rangle^{\alpha=0}\otimes |0\rangle$
with spin
$j={1\over 2}\pm i\sqrt{k-1/4}$.
These states exist for $k\geq 1/4$. In particular, for $k=1/4$ this
is simply the cosmological constant term. But 
these states are not annihilated by any of the four supercharges
$G_{-{1\over 2}},\overline{\tG}_{-{1\over 2}},
\overline{G}_{-{1\over 2}},\tG_{-{1\over 2}}$.
Hence these deformations break supersymmetry and do not concern us.

Finally, as in the bosonic case, we should allow deformations which correspond
to non-normalizable states, if their vertex operators do not grow towards
$\rho\ra\infty$. This means that we should relax the constraint on $j$ for the
discrete series to $0\leq j\leq \frac{k}{2}+1.$ It is easy to check that this
does not yield any new deformations in the supercoset which would preserve
all the symmetries. The only effect of allowing such non-normalizable states
is to extend the range of $k$ for which the operators (\ref{st1})
and (\ref{st2}) exist: if we do not impose normalizability, then they exist
for all $k>0$.

To summarize, in the
$SL(2,\RR)/U(1)$ Kazama-Suzuki model there are two
$(1,1)$ operators that could 
lead to supersymmetric marginal deformations preserving momentum.
They combine into a twisted superpotential term and correspond to the
FI-Theta deformation~(\ref{FItheta}) of the GLSM.
However, the real part is trivial and can be absorbed into a field
redefinition, while the imaginary part is parity-odd and will not be
generated if the high-energy theory is parity even (i.e. has $\theta=0$).
We conclude that the $SL(2,\RR)/U(1)$ Kazama-Suzuki model is rigid,
and does not admit non-trivial deformations preserving all the symmetries.
This in turn implies that the GLSM~(\ref{action}) flows to this superconformal
theory for all $k>0$. 

Note that unlike in bosonic case, there is no puzzle
associated with the absence of a radius-changing operator, because the dual
${\cal N}=2$ Liouville theory does not have it either. While
in the sine-Liouville theory the asymptotic radius and the central charge can 
be varied independently, in the supersymmetric case the radius is quantized
in units of $1/\sqrt k$ if the central charge is $3+6/k$. The easiest
way to see this is to notice that given the action~(\ref{intro:Liouville}), 
one still has
the freedom to choose the period of $\iim\ Y$. The form of the
superpotential constrains the period to be $2\pi n, n\in {\mathbb N}$.
We will see in the next section that the $SL(2,\RR)/U(1)$ supercoset is 
dual to ${\cal N}=2$ Liouville with the smallest possible radius 
corresponding to $n=1$.

\section{Liouville Theory As The Mirror}

In this section, we will find a dual description of our gauge
theory, using the method of \cite{HV}.
We will see that the dual
theory flows in the IR limit to ${\cal N}=2$ supersymmetric
Liouville theory.
Thus we may conclude that the fermionic 2d Black Hole is mirror
to ${\cal N}=2$ Liouville theory.

\subsection{The Dual Theory}

We start with the classical dualization of the system.
We T-dualize the phase of $\Phi$ as well as the imaginary part of $P$.
The dual of a charged chiral superfield
is a neutral twisted chiral superfield which is 
coupled to the field strength superfield $\Sigma$ via a twisted
superpotential. In other words, the lowest component of the dual
superfield is a dynamical theta-angle.
Using the method of~\cite{RV,HV},
we find that the dual action is
\beqa
\widetilde{S}
&=&\frac{1}{2\pi}\int\dd^2x\left\{
\int\dd^4\theta\left[\,-{1\over 2e^2}|\Sigma|^2
-\frac{1}{2}\left(Y+\bY\right)\log\left(Y+\bY\right)-
{1\over 2k}|Y_P|^2\right]\right.
\nonumber\\
&&\left.~~~~~~~~~~~~~
+\frac{1}{2}\left(\int\dd^2\widetilde{\theta}\,\Sigma(Y+Y_P)
+h.c.\right)\right\},
\label{dual1}
\eeqa
where $Y$ and $Y_P$ are the duals of $\Phi$ and $P$ respectively.
The lowest components of both $Y$ and $Y_P$ are periodically
identified with period $2\pi i$.
Gauge-invariant composites of the original fields
are expressed in terms of the dual fields as
\beqa
&&\bPhi\e^V\Phi=\frac{1}{2}\left(Y+\bY\right),\\
&&P+\bP+V=\frac{1}{k}\left(Y_P+\bY_P\right).
\eeqa

Let us now include perturbative quantum corrections to this
dualization procedure. The one-loop divergence in the $|\phi|^2$
one-point function requires an additive renormalization
of $P$:
\beq
P(\Lambda_{\rm UV})=P(\mu)
-\frac{1}{k}\log\left(\frac{\Lambda_{\rm UV}}{\mu}\right).
\eeq
Here $\Lambda_{\rm UV}$ is the UV cut-off.
This induces a similar renormalization of $Y_P$ and hence
of $Y$ so that the twisted F-term in (\ref{dual1}) is finite:
$$
Y(\Lambda_{\rm UV})=Y(\mu)+\log\left({\Lambda_{\rm UV}\over\mu}\right),\qquad
Y_P(\Lambda_{\rm UV})=Y_P(\mu)-\log\left({\Lambda_{\rm UV}\over\mu}\right).
$$
This renormalization does not affect the twisted F-term
but changes the $Y$-part of the D-term in (\ref{dual1}).
In particular, the K\"ahler metric for $y=y(\mu)$ and 
$y_P=y_P(\mu)$ is given by
\beq\label{kahler}
\dd s^2={|\dd y|^2\over 2\log(\Lambda_{UV})+2{\rm Re}\,y}
+{1\over k}|\dd y_P|^2.
\eeq
In the continuum limit $\Lambda_{\rm UV}\ra\infty$, the metric for 
$y$ degenerates to zero, but that for $y_P$ remains finite.

The axial R-rotation shifts the imaginary part of $y$ \cite{HV} as
\beq
{\rm Im}\,Y\ra {\rm Im}\,Y-2\beta.
\eeq
If not for $Y_P$, this anomalous transformation law would induce a change
in the theta-angle. This is a reflection of the fact that 
in the theory of $\Phi$ and $V$ the axial R-symmetry is anomalous.
In the presence of $Y_P$ the anomaly can be cancelled by assigning
an anomalous transformation law to $Y_P$:
\beq
{\rm Im}\,Y_P\ra {\rm Im}\,Y_P+2\beta.
\eeq

Thus the dual system has an axial R-symmetry such that $\e^{-Y}$ and
$\e^{-Y_P}$ have axial R-charges $2$ and $-2$, respectively. The vector
R-charge is zero in both cases.
The anomalous trasnformation law for $Y_P$ corresponds to the modification
$j_A^{\pm}\to j_A^{\pm}\mp 2A_{\mp}$ of
the axial R-current in the original system.

Finally, let us include non-perturbative effects.
The vortex-instanton of the original gauge system can generate
a twisted superpotential in the dual theory.
To find the precise form of the superpotential, it is best to
extend the gauge symmetry to $U(1)_1\times U(1)_2$,
where $U(1)_1$ acts as the phase rotation of $\Phi$
while $U(1)_2$ shifts the imaginary part of $P$.
The first system is the ${\cal N}=2$ QED with one massless flavor,
which has been studied in detail in \cite{HV}. The twisted
superpotential of its dual theory is
\beq
\widetilde{W}_1=\Sigma_1\,Y+\,\e^{-Y}.
\label{inst}
\eeq
The correction term $\e^{-Y}$ is generated by the
vortex-instantons of the $(\Phi,V_1)$ system.
On the other hand,
the system of $P$ and $V_2$ is equivalent to a free theory of a 
massive vector multiplet.
Hence the classical dualization is exact, and the twisted superpotential 
is
\beq
\widetilde{W}_2=\Sigma_2\,Y_P.
\eeq
The absence of the vortex-instanton correction can also be understood by 
noting that the $(P,V_2)$ system has no vortex solutions
because the target space for $P$ is $\RR\times {\mathbb S}^1$.

To get back to the original GLSM~(\ref{action})
we only have to freeze $\Sigma_1-\Sigma_2$ by tuning
the D-term couplings~\cite{HV}. 
Since changing the D-terms cannot affect the twisted F-terms,
the twisted superpotential of
the dual theory is exactly given by
\beq
\widetilde{W}=\Sigma(Y+Y_P)+\e^{-Y}.
\eeq

An alert reader should have noticed that a similar argument
can be used to ``prove'' that the $\oQ_+$-cohomology and hence the
IR central charge of a gauged linear 
sigma-model is independent of the D-terms. On the other hand, we have
seen in Section~\ref{ccharge} that for the GLSM~(\ref{action}) the
$\oQ_+$ cohomology and the central charge do depend on $k$ in a non-trivial
way. In fact, this is crucial for the whole approach described here. 
The loophole in the formal
argument is that it requires integration by parts on the target space of
the low-energy sigma-model. This can be easily seen in the path-integral
formulation. Thus if the target space is noncompact, and the D-term
deformation does not decay fast enough at infinity, then the formal 
argument may fail. Varying $k$ changes the asymptotic behavior of the
target-space metric, and therefore it is not surprising that the 
$\oQ_+$-cohomology depends on $k$. On the other hand, modifying the
gauge couplings has a vanishingly small effect at infinity, because
the gauge fields are massive there.

We have no control over the K\"ahler potential of the dual theory, as
it can get both perturbative and nonperturbative corrections.
The only statement that we can make is that 
the corrections to the semi-classical expression~(\ref{kahler})
are small for ${\rm Re}\,Y\ra +\infty$ and $-{\rm Re}\,Y_P\ra +\infty$, 
because the gauge fields are very massive in this region, and the
interactions are negligible.

\subsection{Liouville theory as the IR limit of the dual theory}

At low energies the vector multiplet $V$, which
has mass of order $e\sqrt{k}$,
can be integrated out.
In the dual theory this gives a constraint
\beq
Y+Y_P=0.
\eeq
Thus we are left with a single twisted chiral superfield
$Y$ with the twisted superpotential
\beq
\widetilde{W}=\e^{-Y}.
\label{Liouville}
\eeq
The above arguments tell us that the K\"ahler 
potential has the form
\beq
K(Y,\bY)=-{1\over 2k}|Y|^2+\ldots,
\label{KL}
\eeq
where the terms denoted by dots go to zero for $\rre\,Y\ra +\infty$.
Otherwise the K\"ahler potential is undetermined.

The superpotential (\ref{Liouville}) is the Liouville potential.
It is known that the theory with this superpotential and a flat
K\"ahler potential $K_{\gamma}=-{1\over 2\gamma^2}|Y|^2$
is a $(2,2)$ superconformal field theory with central charge
\beq
c=3\left(1+{2\over \gamma^2}\right).
\eeq
In fact the current superfield
\beq
\widetilde{\cJ}={1\over 2\gamma^2}\bD_-YD_-\overline{Y}
+{1\over \gamma^2}(\partial_0-\partial_1){\rm Im}\,Y
\label{tiJ}
\eeq
obeys $\bD_+\widetilde{\cJ}=0,$ and the lowest components
of $\widetilde{\cJ},D_-\widetilde{\cJ},\bD_-\widetilde{\cJ},
{1\over 4}[\bD_-,D_-]\widetilde{\cJ}$ generate ${\cal N}=2$
superconformal algebra with central charge $c=3+6/\gamma^2$.
The linear term in (\ref{tiJ}) shows that there is a linear dilaton with
the slope proportional to $1/\gamma^2$.

We are now very close to proving that the IR limit of the dual theory
is the ${\cal N}=2$ Liouville theory with $\gamma^2=k$. Indeed, we already know
that the twisted superpotential, the central charge, and
the asymptotic behavior of the K\"ahler potential are the same in the two
theories, if we set $\gamma^2=k$. But there still remains a remote
possibility that there is another twisted Landau-Ginzburg model with the 
same central charge and twisted superpotential, but different K\"ahler
potential, which nevertheless has the same asymptotics. One could rule out
the existence of  such a ``fake'' Liouville theory in the neighborhood
of the ordinary Liouville theory by studying marginal deformations of
the latter. We take an alternative route, which directly demonstrates
that the dual of the GLSM flows to the ${\cal N}=2$ Liouville theory.
Let us dualize the phase of $P$ only, leaving $\Phi$ as it is. As explained
above, the classical dualization is exact in this case. The resulting 
gauged linear sigma-model has both twisted and ordinary chiral fields
and the following action
\beq\label{dualaction}
S={1\over 2\pi}\int\dd^2x\, \dd^4\theta\left[
\bPhi \e^V\Phi-\frac{1}{2k}|Y_P|^2-{1\over 2e^2}|\Sigma|^2\right]+
{1\over 4\pi}\left(\int \dd^2x\,\dd^2\widetilde{\theta}\, \Sigma Y_P + h.c.\right).
\eeq
Recall now that the twisted chiral superfield $e^{Y_P}$ has axial 
R-charge $2$ and vector R-charge $0$. Hence we can deform the above theory 
by adding a twisted superpotential 
\beq\label{addtwist}
\frac{\kappa}{4\pi}\int\dd^2x\,\dd^2\widetilde{\theta}\, e^{Y_P}+h.c.
\eeq
without breaking the axial R-symmetry. It follows that this deformation
results in an exactly marginal deformation of the IR fixed point which
does not change the central charge and preserves $(2,2)$ supersymmetry. 
Furthermore, the asymptotic region in the undeformed theory
corresponds to $|\Phi|\ra\infty, \rre\,Y_P\ra -\infty.$
Since the twisted superpotential~(\ref{addtwist}) is exponentially
small in this region, this deformation does not change the asymptotic
behavior of the model. Note
also that after the twisted superpotential has been added, we cannot
dualize back to the $(\Phi,P,V)$ variables.

Now we recall that the fermionic 2d Black Hole does not have
non-trivial marginal deformations preserving $(2,2)$ supersymmetry. 
It follows that the model~(\ref{dualaction}) deformed by the twisted 
superpotential~(\ref{addtwist}) flows to the fermionic 2d black
hole at level $k$ for all $\kappa$. We can use this to our advantage
by taking the limit $\kappa\ra\infty$. To see what happens in this limit,
we set $Y_P=\tY_P-\log(\kappa/\kappa_0)$ so that in terms of $\tY_P$ the
twisted superpotential remains fixed. In terms of $\Phi,\tY_P$ and $V$
the action becomes
\beqa\label{deformedaction}
S&=&{1\over 2\pi}\int\dd^2x\,\dd^4\theta\left[
\bPhi \e^V\Phi-\frac{1}{2k}|\tY_P|^2-{1\over 2e^2}|\Sigma|^2\right]\\
&+&
\left({1\over 4\pi}\int\dd^2x\,\dd^2\widetilde{\theta}\, \left[\Sigma 
\left(-\log\frac{\kappa}{\kappa_0} + \tY_P\right)+\kappa_0 e^{\tY_P}\right]+
h.c.\right).
\eeqa
We see that $\rre\log (\kappa/\kappa_0)$ plays the role of the Fayet-Iliopoulos
term. For $\kappa\ra\infty$ the Fayet-Iliopoulos term breaks the gauge 
symmetry at a very high scale of order $\log\kappa$, the gauge field
eats $\Phi$, and all the fields except $\tY_P$ get a mass of order 
$\log\kappa.$ 
Integrating them out classically, we are left with a twisted chiral 
superfield $\tY_P$ with a twisted superpotential $e^{\tY_P}$ and a 
K\"ahler potential
$$
-\frac{1}{2k}|\tY_P|^2
$$
This is an ${\cal N}=2$ Liouville theory with central charge $c=3+6/k.$
As $\kappa$ increases, the accuracy of the classical approximation becomes
arbitrarily good. On the other hand, we know that the IR limit of the
theory does not depend on $\kappa$ at all. Hence the GLSM flows to the 
${\cal N}=2$ Liouville theory for all $\kappa$, including $\kappa=0$. 
This concludes the argument.

\section{Some Generalizations}

In this section, we discuss a few generalizations of our setup.
One generalization is to consider an orbifold of the fermionic 2d Black 
Hole background with respect to a discrete subgroup of the $U(1)$ 
isometry.\footnote{This idea arose in a conversation with Juan Maldacena.}
Other generalizations are sigma-models on higher dimensional manifolds,
some of which can be used to construct dilatonic
superstring backgrounds, while others have a mass gap.

\subsection{Orbifolds}

In ${\cal N}=2$ Liouville theory (\ref{intro:Liouville}),
the form of the superpotential $\e^{-Y}$ constrains
 the periodicity of $\iim\, Y$ to be an integer multiple of $2\pi$,
and therefore the radius of the circle parametrized by $\iim\, Y$
is quantized  in units of $1/\sqrt k$.
As mentioned in Section~\ref{flowtwo},
this is an important difference between the
${\cal N}=2$ Liouville theory and its bosonic relative,
the sine-Liouville theory:
in the latter the radius of the circle can be varied independently of $k$.
We have shown that the ${\cal N}=2$ Liouville theory with the minimal radius 
$1/\sqrt k$ is mirror to the $SL(2,\RR)/U(1)$ supercoset. What about the other 
values of the radius? Since asymptotically mirror transformation reduces
to T-duality, the mirror for Liouville theory with radius $n/\sqrt k$ must
be some generalization of the supercoset with asymptotic radius $\sqrt k/n$.
An obvious guess is an orbifold of the
supercoset by a $\ZZ_n$ subgroup of the momentum symmetry. 

To show that this guess is correct, note that the 
orbifoldized supercoset can be obtained by orbifoldizing the 
GLSM~(\ref{action}) by the same symmetry. This means that one should
take the period of $\iim\ P$ to be $2\pi/n$ instead of $2\pi$. To derive the
mirror of such a model, we use the approach explained in subsection 6.2:
we T-dualize $P$ to a twisted chiral multiplet $Y_P$ and add a twisted
superpotential $e^{Y_P}$. As we increase the coefficient of $e^{Y_P}$,
the theory is smoothly deformed to ${\cal N}=2$ Liouville theory. The only
difference is that the period of $\iim\ Y_P$ is now $2\pi n$ instead of 
$2\pi$. This proves that the $\ZZ_n$ orbifold of the supercoset is mirror
to the ${\cal N}=2$ Liouville theory with radius $n/\sqrt k$.

Note that the $\ZZ_n$ action on the fermionic 2d Black Hole has a 
fixed point at the tip of the cigar. Thus the orbifoldized sigma-model
metric has a conical singularity with a deficit angle $2\pi(1-1/n)$. 
Nevertheless the conformal field theory
is well-defined. In the bosonic case, FZZ duality suggests that the
cigar with an arbitrary conical deficit leads to a well-defined
CFT, but it is not known how to see this directly.

\subsection{Multi-variable Models}

It is straightforward to generalize the story to theories
with a larger number of fields and higher rank gauge groups.
Let us consider a $U(1)^M$ gauge theory with $N+M$ matter fields
$\Phi_i$ ($i=1,\ldots,N$), $P_{\ell}$ ($\ell=1,\ldots,M$)
where the gauge transformation is defined by
$\Phi_i\to\e^{i\sum_{\ell=1}^MR_{i\ell}\Lambda_{\ell}}\Phi_i$ and
$P_{\ell}\to P_{\ell}+i\Lambda_{\ell}$.
The fields $P_{\ell}$ are periodic in the imaginary direction and
we take all periodicities to be $2\pi i$.
The action of the system is given by
\beq
S={1\over 2\pi}\int\dd^2x\,\dd^4\theta\left[
\sum_{i=1}^N\bPhi_i \e^{R_i\cdot V}\Phi_i
+\sum_{\ell=1}^M{k_{\ell}\over 4}(P_{\ell}+\bP_{\ell}+V_{\ell})^2
-\sum_{\ell=1}^M{1\over 2e_{\ell}^2}|\Sigma_{\ell}|^2\right],
\label{action2}
\eeq
where $R_i\cdot V=\sum_{\ell=1}^MR_{i\ell}V_{\ell}$.
The chiral anomaly equation
$\partial_{\mu}j_A^{\mu}=2\sum_{i=1}^NR_i\cdot F_{+-}$
has a supersymmetric extension 
\beq
\bD_+\cJ^{\circ}={1\over 2}\sum_{i,\ell}R_{i\ell}\bD_-\Sigma_{\ell},
\eeq
where $\cJ^{\circ}$ is defined by
\beqa
\cJ^{\circ}&=
&\sum_{i=1}^ND_-(\Phi_i\e^{R_i\cdot V})\e^{-R_i\cdot V}
\bD_-(\e^{R_i\cdot V}\bPhi_i)
\nonumber\\[-0.1cm]
&&+\sum_{\ell=1}^M\left\{\,
{k_{\ell}\over 2}D_-(P_{\ell}+\bP_{\ell}+V_{\ell})
\bD_-(P_{\ell}+\bP_{\ell}+V_{\ell})
+{i\over 2e_{\ell}^2}\Sigma_{\ell}(\partial_0-\partial_1)
\bSigma_{\ell}\right\}.
~~~~~~~~~
\eeqa
The modified current
\beq
\cJ=\cJ^{\circ}+{1\over 2}\sum_{i,\ell}
R_{i\ell}[\bD_-,D_-](P_{\ell}+\bP_{\ell}+V_{\ell})
\label{modicu}
\eeq
satisfies the right-chiral condition $\bD_+\cJ=0$.
The components of this current form an 
${\cal N}=2$ superconformal algebra with central charge
\beq
c=3\left(N+\sum_{\ell=1}^M{2b_{\ell}^2\over k_{\ell}}\right),
\label{cen}
\eeq
where $b_{\ell}:=\sum_{i=1}^NR_{i\ell}$.
If we make a natural assumption that for large
$-{\rm Re}\,P_{\ell}$
the theory flows to a free theory, we can argue as before that
the full theory flows to a SCFT with central charge given by
(\ref{cen}).
The linear terms in (\ref{modicu}) show that
there is a linear dilaton in such an asymptotic region,
with the components of the gradient proportional to the chiral anomaly
coefficients $b_{\ell}$.
Thus the system describes a $2N$-dimensional background with a non-trivial
dilaton profile (except in the case where all $b_{\ell}$ vanish).

As before, dualization of $\Phi_i$ and $P_{\ell}$ turns them into twisted
chiral superfields $Y_i$ and $Y_{P_{\ell}}$ of period $2\pi i$.
The twisted superpotential is given by
\beq
\widetilde{W}=\sum_{\ell=1}^M\Sigma_{\ell}
\left(\sum_{i=1}^NR_{i\ell}Y_i+Y_{P_{\ell}}\right)
+\sum_{i=1}^N\e^{-Y_i},
\eeq
where the exponential terms are from $\Phi_i$ vortices.
The K\"ahler potential for $Y_i$ is vanishingly small in the continuum limit,
but that for
$Y_{P_{\ell}}$ remains finite and equal to 
$$
-|Y_{P_{\ell}}|^2/2k_{\ell}.
$$
In the infrared limit $e_{\ell}\to\infty$, it is appropriate to integrate out
the gauge multiplets, which imposes a constraint
$\sum_{i=1}^NR_{i\ell}Y_i+Y_{P_{\ell}}=0$.
Thus we are left with a theory of $N$ fields $Y_i$
with the following K\"ahler potential and superpotential:
\beqa
&&K=-{1\over 2}\sum_{i,j=1}^N g_{ij}\overline{Y}_iY_{j}
+\ldots,
\label{Kah}\\
&&\widetilde{W}=\sum_{i=1}^N\e^{-Y_i},
\label{Tsup}
\eeqa
where the terms denoted by dots are small in the asymptotic region, and 
$g_{ij}$ is given by
\beq
g_{ij}=\sum_{\ell=1}^MR_{i\ell}{1\over k_{\ell}}R_{j\ell}.
\label{mij}
\eeq
If we omit the terms denoted by dots in the K\"ahler potential,
then the theory is conformally
invariant, with the superconformal algebra generated by the
supercurrent
\beq
\widetilde{\cJ}=\sum_{i,j}{1\over 2}g_{ij}D_-Y_i\bD_-\overline{Y}_{j}
+\sum_{ij}g_{ij}(\partial_0-\partial_1){\rm Im}\,Y_{j}.
\eeq
Its central charge is given by (\ref{cen}). This suggests that
the terms denoted by dots in (\ref{Kah}) vanish in the IR limit.

\subsection{Squashed Toric Sigma-Models}

Including matter fields transforming
inhomogeneously under the gauge group, like $P$ in our theory,
provides interesting generalizations of the standard linear sigma-models.
In this way one can obtain not
only new superconformal field theories, but also new massive ${\cal N}=2$
field theories.
Since this topic is beyond scope of this paper,
we shall only briefly comment on it.

Consider a $U(1)^k$ gauge theory with $N$ chiral superfields $\Phi_i$
of charge $Q_i^a$ ($i=1,\ldots,N$, $a=1,\ldots,k$) and
FI-Theta parameters $t^a=r^a-i\theta^a$.
It has a flavor symmetry group $U(1)^{N-k}$ acting on
$\Phi_i$ with charges $R_{i\ell}$ ($\ell=1,\ldots N-k$) complementary to
$Q_i^a$.
For a suitable choice of $r^a$,
the space of classical vacua is a toric manifold $X$ of dimension $N-k$
where the $U(1)^{N-k}$ flavor group action
determines the structure of the torus fibration.
The metric on $X$ is obtained by the standard K\"ahler reduction.
For example, for $U(1)$ gauge theory with two charge 1 chiral fields
the classical moduli space is $X=\CP^1$ with the round (Fubini-Study) metric. 
At low energies the theory reduces  to the non-linear supersymmetric sigma
model on $X$.

Now let us consider the following deformation of this system.
We gauge the $U(1)^{N-k}$ flavor group and introduce
for each $U(1)$ factor
a chiral superfield $P_{\ell}$ transforming inhomogeneously.
The action of the system reads
\beqa
S&=&{1\over 2\pi}\int\dd^2x\,\left\{\int\dd^4\theta\Biggl[~
\sum_{i=1}^N\bPhi_i \e^{Q_i\cdot V+R_i\cdot V'}\Phi_i
-\sum_{a=1}^{k}
{1\over 2e_{a}^2}|\Sigma_{a}|^2\Biggr]
+{\rm Re}\int \dd^2\widetilde{\theta}\sum_{a=1}^kt^a\Sigma_a
\right.
\nonumber\\
&&~~~~~~~~~~~~~~~~~~
\left.+\int\dd^4\theta\Biggl[\,
\sum_{\ell=1}^{N-k}{k_{\ell}\over 4}(P_{\ell}+\bP_{\ell}+V'_{\ell})^2
-\sum_{\ell=1}^{N-k}{1\over 2e_{\ell}^2}|\Sigma'_{\ell}|^2
~\Biggr]\right\},~~~~
\label{action3}
\eeqa
where $Q_i\cdot V=\sum_{a=1}^kQ_i^aV_a$ and
$R_i\cdot V'=\sum_{\ell=1}^{N-k}R_{i\ell}V'_{\ell}$.
The vacuum manifold $X'$ is
again a toric manifold with the same complex structure and the same
K\"ahler class as $X$, but with a different K\"ahler metric.
For large $r^a$'s,
deep in the interior of the base of the torus fibration,
the sizes of the torus fibers
are constants proportional to $\sqrt{k_{\ell}}$. We will say that
$X'$ is a ``squashed version'' of the toric manifold,
and we obtain the sigma model on a squashed toric manifold
at low energies.
For $X=\CP^1$ (round 2-sphere), $X'$ looks like a sausage,
so we obtain a supersymmetric version of the
``sausage model'' of \cite{sausage}.
In the limit $k_{\ell}\to\infty$,
the $P_{\ell}$-$\Sigma'_{\ell}$ pairs decouple, and
we recover the sigma-model on the ``round toric manifold'' $X$.

The theory is expected to flow to a non-trivial superconformal field
theory when $\sum_{i=1}^NQ_i^a=0$ for all $a$.
If this condition is fulfilled, then the central charge of the IR fixed point
is
\beq
c=3\left(N-k+\sum_{\ell=1}^{N-k}\frac{2b_{\ell}^2}{k_{\ell}}\right),
\eeq
where $b_{\ell}=\sum_{i=1}^{N}R_{i\ell}$.
In the limit $k_{\ell}\to\infty$ (no squashing),
$c/3$ becomes the complex dimension $N-k$ of the manifold $X$.

The dual theory is found as above, i.e. by dualizing $\Phi_i$ and $P_{\ell}$,
taking account of the $\Phi_i$-vortices, and
integrating out the gauge multiplets.
We find that the dual K\"ahler potential and twisted superpotential are
\beq
K=-{1\over 2}\sum_{i,j=1}^N
g_{ij}\overline{Y}_iY_j+\ldots,~~~
\widetilde{W}=
\sum_{i=1}^N\e^{-Y_i},
\eeq
where $g_{ij}$ is defined by (\ref{mij}) with $M=N-k$.
This time, however,
integration over the gauge multiplets $\Sigma_a$ imposes
a constraint
\beq
\sum_{i=1}^NQ_i^aY_i=t^a.
\eeq
This is the mirror of the sigma-model on the squashed toric manifold $X'$.
It is the same as the mirror of the sigma-model on the ``round toric''
$X$, except that the K\"ahler potential is now finite
whereas that for the mirror of $X$ is vanishingly small in the continuum
limit~\cite{HV}.
For example, when $X=\CP^1$, $X'$ is sausage-shaped, and
we find that the mirror of the supersymmetric sausage
model is the ${\cal N}=2$ sine-Gordon model with a finite K\"ahler potential.
This equivalence has been conjectured by Fendley and Intriligator~\cite{FIpr}
as a natural generalization of~\cite{FeIn}.

The introduction of matter fields which
transform inhomogeneously under the gauge group
is analogous to the introduction of ``magnetic'' gauge fields
with BF couplings
in $2+1$ dimensional gauge theories~\cite{KaSt}. 
In fact, in $2+1$ dimensions they are related by abelian electric-magnetic
duality. Mirror symmetry between a squashed toric sigma-model
and the Landau-Ginzburg model with a finite K\"ahler potential can also be 
derived from the all-scale ${\cal N}=4$ mirror symmetry in $2+1$ 
dimensions~\cite{KaSt} by an RG flow to an ${\cal N}=2$ mirror~\cite{DT} and
further compactification to $1+1$ dimensions~\cite{AHKT}.

\section{Concluding Remarks}

We have proved the equivalence of the $SL(2,\RR)/U(1)$ Kazama-Suzuki 
supercoset model and ${\cal N}=2$ Liouville theory.
We first argued that a super-renormalizable gauge theory
flows to the $SL(2,\RR)/U(1)$ supercoset model.
The argument had three ingredients: the analysis of the
RG flow in the one-loop approximation which is valid for $k\gg 1$,
an exact computation of the infrared central charge of the gauge theory,
and the analysis of marginal deformations of the supercoset.
We then used the argument of~\cite{HV} to find the dual description
of the gauge theory. This dual theory flows in the IR limit to
the ${\cal N}=2$ Liouville theory. We also gave an alternative derivation
of the mirror dual by showing that the gauge theory can be continuosly
deformed to the ${\cal N}=2$ Liouville theory while leaving the infrared fixed
point unchanged. 

This example teaches us an important lesson:
a super-renormalizable gauge theory can flow to a background with 
a non-trivial 
dilaton profile, including a region with a linear dilaton.
We have shown how the dilaton is generated in two different ways: 
by a one-loop analysis of the RG 
flow, and by computing the currents of the ${\cal N}=2$ superconformal
algebra in the topologically twisted gauge theory. In the first instance,
we observed that the RG flow has a well-defined fixed point
only if the target-space coordinates transform non-trivially under the
Weyl rescaling. This is a signature of the dilaton gradient.
In the second instance, we saw that na{\"\i}ve 
superconformal generators must be corrected by terms linear in fields
because of the axial/Konishi anomaly of the gauge theory.

Another interesting aspect of this work is that it sheds some light
on the relation between coset models and Landau-Ginzburg 
models.
It has been known for a long time that many (super)conformal field 
theories can be realized as coset models as well as the IR limits of 
Landau-Ginzburg models,
but the relation between the two descriptions has not been well understood.
The present work is the first example where the two descriptions
are connected in a rather transparent way.
It would be interesting to see if the methods of this paper can be
extended to other models,
for example, the $K$-th ${\cal N}=2$ unitary minimal model
which can be realized as the $SU(2)_K/U(1)$ Kazama-Suzuki model
or as the IR limit of the Landau-Ginzburg model with the superpotential $W=X^{K+2}$.
In fact, the equivalence of the two models motivated
the observation of~\cite{MV,GM}
that certain correlation functions of the
$SL(2,\RR)_{1+2}/U(1)$ Kazama-Suzuki model
and the $W=X^{-1}$ Landau-Ginzburg model agree.
(As pointed out in the first reference of~\cite{Nmatrix},
for certain purposes $SL(2,\RR)_K/U(1)$ can be regarded as an
analytic continuation of $SU(2)_K/U(1)$ to negative $K$.)
More generally, it was proposed in~\cite{OV} that there is a relation
between the $SL(2,\RR)_{k+2}/U(1)$ Kazama-Suzuki model and 
$(W=X^{-k})/\ZZ_k$ Landau-Ginzburg orbifold (for integer $k$).
As should be clear by now,
these observations and conjectures can be regarded as a consequence of
the supersymmetric FZZ duality in the special case $k\in {\mathbb N}$,
if we identify $\e^{-Y}$ with
$X^{-k}$.

Our research was partly motivated by the bosonic FZZ duality.
In this paper, we have only considered the supersymmetric version,
but it is important to understand the FZZ duality itself.
One could attempt to apply the methods of this paper to gain some 
understanding of this duality. For example, one could try to find
a super-renormalizable gauge theory which flows to the bosonic coset model,
and then look for a dual description.
Without supersymmetry, one may not be able to make an exact statement,
but one may be able to see qualitatively how the FZZ duality emerges.
Alternatively, one could start with the supersymmetric FZZ duality
and consider a supersymmetry breaking perturbation which is
relevant or marginally relevant and gives a mass to the fermions but not to 
the bosons.
Then one should analyze the corresponding perturbation of the
${\cal N}=2$ Liouville theory.
In particular, it would be interesting to understand the origin of the 
restriction $k>2$ in the bosonic FZZ duality.

Another interesting direction to pursue is to study D-branes 
in the supercoset/Liouville theory.
Since this SCFT is relevant for both the deformed conifold
and the ALE space~\cite{GV,OV,GK1,ES1,MSNN}, such a study
should improve our understanding of D-brane dynamics near the 
conifold and ALE singularities.
The supercoset/Liouville theory also describes Little String Theories
in a double scaling limit~\cite{GK1}, so there should also be a connection
with D-branes in the presence of NS 5-branes.
For a discussion of D-branes in ${\cal N}=2$ Liouville theory
and for references on D-branes in a linear dilaton background,
see for example~\cite{EgSu}.
The relation between the descriptions of
D-branes in the coset models and in the Landau-Ginzburg models
also deserves study, and the present work may be
useful in this regard.

\appendix{Conventions}
\label{app:A}

Here we record our conventions for superfields
on $(2,2)$ superspace with coordinates
$x^0,x^1$ (bosonic),
$\theta^+,\theta^-,\btheta^+,\btheta^-$ (fermionic).
The bosonic coordinates span the flat Minkow\-ski space
(metric $\eta_{00}=-1$,
$\eta_{11}=1,$ and $\eta_{01}=0$)
and we often use the light cone coordinates $x^{\pm}=x^0\pm x^1$
and derivatives
$\partial_{\pm}:=\partial/\partial x^{\pm}=(\partial_0\pm\partial_1)/2$.
The fermionic coordinates are related
by complex conjugation: $(\theta^{\pm})^\dagger=\btheta^{\pm}$.

Supersymmetry transformation is represented on superfields
by derivative operators
\bea
{\cal Q}_{\pm}&=&\frac{\partial}{\partial \theta^{\pm}}
+i \overline{\theta}^{\pm}\,\partial_{\pm},
\label{Qpm}\\
\overline{{\cal Q}}_{\pm}
&=&
-\frac{\partial}{\partial \overline{\theta}^{\pm}}
-i \theta^{\pm}\,\partial_{\pm},
\label{oQpm}
\eea
which obey
$\{ {\cal Q}_{\pm},\overline{{\cal Q}}_{\pm}\} =-2 i \partial_{\pm}$.
Another pair of derivatives
\bea
D_{\pm}&=&\frac{\partial}{\partial \theta^{\pm}}
-i \overline{\theta}^{\pm}\,\partial_{\pm}, \\
\overline{D}_{\pm}&=&-\frac{\partial}{\partial \overline{\theta}^{\pm}}
+i \theta^{\pm}\,\partial_{\pm},
\eea
anti-commutes with ${\cal Q}_{\pm}$, $\overline{{\cal Q}}_{\pm}$,
and obeys
$\{D_{\pm},\overline{D}_{\pm}\}=2i\partial_{\pm}$.
Vector/axial R-rotations are
\bea
\e^{i\alpha F_{V}}:{\cal F}(x^{\mu},\theta^{\pm},\btheta^{\pm})\mapsto
\e^{i \alpha q_{V}}
{\cal F}(x^{\mu},\e^{-i\alpha}\theta^{\pm},
\e^{i\alpha}\overline{\theta}^{\pm})\\
\e^{i\beta F_{A}}:{\cal F}(x^{\mu},\theta^{\pm},\btheta^{\pm})\mapsto
\e^{i \beta q_{A}}{\cal F}(x^{\mu},\e^{\mp i\beta}\theta^{\pm},
\e^{\pm i\beta}\overline{\theta}^{\pm}),
\eea
where $q_V$/$q_A$ are the vector/axial R-charges of ${\cal F}$.
A chiral superfield $\Phi$ obeys
$\overline{D}_{\pm}\Phi =0$, while
a twisted chiral superfield $U$ obeys
$\overline{D}_+U=D_-U=0$.
A supersymmetric action is constructed from D-terms, F-terms, and
twisted F-terms which are given by the following superspace
integrals respectively:
\beqa
&&\int \dd^{2}x \,\dd^{4}\theta \,
K({\cal F}_i)=\int \dd^2x \,\dd\theta^+\dd\theta^-\dd\btheta^-\dd\btheta^+\,
K({\cal F}_i),
\label{Dterm}\\
&&\int \dd^{2}x \dd^{2}\theta\, W(\Phi_i)
=\int \dd^2x\, \dd\theta^- \dd\theta^+\,
W(\Phi_i)\Bigr|_{\btheta^{\pm}=0},
\label{Fterm}\\
&&\int \dd^2x\,\dd^2\widetilde{\theta}\,\widetilde{W}(U_i)
=\int \dd^2x\,\dd\btheta^-\dd\theta^+\,\widetilde{W}(U_i)
\Bigr|_{\btheta^+=\theta^-=0}.
\label{tFterm}
\eeqa
Here $K(-)$ is an arbitrary differentiable function
of arbitrary superfields ${\cal F}_i$,
$W(\Phi_i)$ is a holomorphic function of chiral superfields
$\Phi_i$,
and $\widetilde{W}(U_i)$ is a holomorphic function of
twisted chiral superfields $U_i$.

The vector superfield in the Wess-Zumino gauge is expressed as
\beqa
V&=&\theta^-\btheta^-(v_0-v_1)+\theta^+\btheta^+(v_0+v_1)
-\theta^-\btheta^+\sigma-\theta^+\btheta^-\bsigma
\nonumber\\
&&
+i\theta^-\theta^+(\btheta^-\blambda_-+\btheta^+\blambda_+)
+i\btheta^+\btheta^-(\theta^-\lambda_-+\theta^+\lambda_+)
+\theta^-\theta^+\btheta^+\btheta^- D.
\label{WZgauge}
\eeqa
The field-strength superfield is given by
\beq
\Sigma:=\overline{D}_+D_-V
=\sigma+i\theta^+\blambda_+
-i\btheta^-\lambda_-
+\theta^+\btheta^-(D-iv_{01})+\cdots,
\eeq
where $v_{01}=\partial_0v_1-\partial_1v_0$.

Let us also fix a convention for the normalization of the
sigma-model action.
For a target space with metric $g_{IJ}$ the sigma-model action
on the two-dimensional Minkowski space will be
\beq
S={1\over 4\pi}\int g_{IJ}
(\partial_0X^I\partial_0X^J-\partial_1X^I\partial_1X^J)\dd^2x.
\eeq

\appendix{OPE of Elementary Fields}
\label{app:B}

In this Appendix we compute the short distance singularity of the
product of two elementary fields of the GLSM~(\ref{action}),
or (\ref{actioncomp}).

The leading singularity for the matter fields is the standard one:
\beqa
&&\phi(x)\bphi(0)\sim -{1\over 2}\log(x^2),
~~~\psi_{\pm}(x)\bpsi_{\pm}(0)\sim {-i\over x^{\pm}},\\
&&p(x)\bp(0)\sim -{1\over k}\log(x^2),
~~~\chi_{\pm}(x)\bchi_{\pm}(0)\sim {-2i/k\over x^{\pm}},\\
&&\sigma(x)\bsigma(0)\sim -e^2\log(x^2),
~~~\lambda_{\pm}(x)\blambda_{\pm}(0)\sim {-2ie^2\over x^{\pm}}.
\eeqa
More subtle is the subleading singularity and the OPE of gauge fields.
To compute them we need to fix the gauge symmetry.
We choose the standard Lorentz gauge. Namely, we add to the action
(\ref{actioncomp}) the term
\beq
-{1\over 2\pi}\int \dd^2x {1\over 8\alpha}(\partial^{\mu}v_{\mu})^2,
\eeq
where $\alpha$ is the gauge parameter that should not
appear in any gauge-invariant physical observables.
Then it is straightforward to derive the following OPE
($\vp:={\rm Im}\,p$)
\beqa
&&\partial_{\pm}\vp(x)\,\partial_{\pm}\vp(0)
\sim
-{1\over 2k}{1\over (x^{\pm})^2}+{\alpha\over 2}{x^{\mp}\over x^{\pm}},\\
&&\partial_+\vp(x)\,\partial_-\vp(0)
\sim {\pi i\over 2k}\delta(x)+{\alpha\over 2}\log (x^2),
\label{ppOPE}\\
&&v_{\pm}(x)\,v_{\pm}(0)\sim
\left({\alpha\over 2}-{e^2\over 8}\right){x^{\mp}\over x^{\pm}},\\
&&
v_+(x)\,v_-(0)\sim \left({\alpha\over 2}+{e^2\over 8}\right)\log (x^2),\\
&&
\vp(x)v_{\pm}(0)\sim -{\alpha\over 2}\partial_{\mp}^{-1}\log(x^2).
\eeqa
{}From this we see that the gauge-invariant current
$A_{\pm}:=\partial_{\pm}\vp+v_{\pm}$
has the following OPE:
\beqa
&&A_{\pm}(x)A_{\pm}(0)\sim -{1\over 2k}{1\over (x^{\pm})^2}
-{e^2\over 8}{x^{\mp}\over x^{\pm}},\\
&&A_+(x)A_-(0)\sim {\pi i\over 2k}\delta(x)+{e^2\over 8}\log(x^2).
\label{AAOPE}
\eeqa
In this paper, we do not use
the equations (\ref{ppOPE}) and (\ref{AAOPE}) that
include delta-functions,
which are convention-dependent contact terms.

\appendix{Konishi Anomaly}
\label{app:C}

Let us define
$:\!\psi_{\mp}(x_1)\opsi_{\mp}(x_2)\!:$ by
$\psi_{\mp}(x_1)\opsi_{\mp}(x_2)-{-i\over x_1^{\mp}-x_2^{\mp}}$.
By a one-loop computation, we find
\beq
\langle \,:\!\psi_-(x_1)\opsi_-(x_2)\!:{\cal O}\,\rangle
\sim
-{i\over \pi}\int{\dd^2z\over(x_1^--z^-)(x_2^--z^-)}
\langle v_+(z){\cal O}\rangle.
\eeq
In particular, we have
\beqa
&&\langle\,:\!\!\!\,(\partial_+\psi_-(x_1)\opsi_-(x_2)
+\psi_-(x_1)\partial_+\opsi_-(x_2))\!\!:{\cal O}\,\rangle
\sim
-\left\langle{v_+(x_1)-v_+(x_2)\over x_1^--x_2^-}\,
{\cal O}\right\rangle\nonumber\\
&&~~~~~~~~~~~~~
\sim
-\left\langle \left\{\partial_-v_+(x_2)+{x_1^+-x_2^+\over x_1^--x_2^-}
\partial_+v_+(x_2)\right\}{\cal O}\right\rangle.
\eeqa
We see that the limit $x_1\to x_2$ is ambiguous.
This ambiguity is absent for the gauge-invariant
current $\psi_-\opsi_-$ defined as 
\beqa
\psi_-\opsi_-(x)&:=&\lim_{x_1\to x_2}
\left(\,\psi_-(x_1)\exp\left(\mbox{$i\int_{x_2}^{x_1}v$}\right)\opsi_-(x_2)
-{-i\over x_1^--x_2^-}\,\right)
\label{gaugeinv}\\[0.1cm]
&=&:\!\psi_-(x)\opsi_-(x)\!:+v_-(x)+\lim_{x_1\to x_2}
{x_1^+-x_2^+\over x_1^--x_2^-}v_+(x).
\label{gaugeinv2}
\eeqa
Indeed, we see that
\beq
\langle \partial_+(\psi_-\opsi_-)(x){\cal O}\rangle
\sim\langle F_{+-}(x){\cal O}\rangle.
\label{pmO}
\eeq
Similarly, if we define
$\psi_+\opsi_+$ as the limit of
$\psi_+(x_1)\exp(i\int_{x_2}^{x_1}v)\opsi_+(x_2)-{-i\over x_1^+-x_2^+},$
we find
\beq
\langle \partial_-(\psi_+\opsi_+)(x){\cal O}\rangle
\sim-\langle F_{+-}(x){\cal O}\rangle.
\label{ppO}
\eeq

Gauge-invariant composites that appears
in the text are always defined by a formula like (\ref{gaugeinv}).
For instance, let us look
at the axial current $j_A^+=\psi_-\opsi_-+\ldots$
and $j_A^-=-\psi_+\opsi_++\ldots$ in (\ref{axial}).
The OPE (\ref{pmO}), (\ref{ppO})
based on such a definition is consistent with
the chiral anomaly equation
$\partial_+j_A^++\partial_-j_A^-=2F_{+-}$.

Such quantum effects can
modify the classical equation
\beq
\bD_+\cJ^{\circ}\stackrel{?}{=}0
\label{DcJ}
\eeq
where $\cJ^{\circ}$ is the superfield
defined in (\ref{Jcirc}).
Let us look at the lowest component of $\cJ^{\circ}$
\beq
j^{\circ}_-=\psi_-\opsi_-+{k\over 2}\chi_-\bchi_-
+{i\over e^2}\sigma\partial_-\bsigma.
\eeq
The equation (\ref{DcJ}) would tell us that it commutes
with $\oQ_+$.
However, when $\psi_-\opsi_-$ is defined as in (\ref{gaugeinv2}),
the commutator becomes
\beq
\left[\oQ_+, j^{\circ}_-\right]={i\over 2}\lambda_-,
\label{Qj}
\eeq
where we have used $[\oQ_+, v_-]={i\over 2}\lambda_-$
and $[\oQ_+,v_+]=0$.
The right-hand side of (\ref{Qj})
is the lowest component of the superfield ${1\over 2}\bD_-\Sigma$.
Hence the supersymmetric completion of (\ref{Qj}) is
\beq
\bD_+\cJ^{\circ}={1\over 2}\bD_-\Sigma,
\label{DJ}
\eeq
which can be regarded as the anomalous form of (\ref{DcJ}).
One can also explicitly check other components of the superfield
equation (\ref{DJ}).
For instance, the $\theta^-$-component equation
$\{\oQ_+,G_-^{\circ}\}=-i\partial_-\sigma$
follows from a one-loop computation,
while the $\theta^-\btheta^-$-component
$\{\oQ_+,T^{\circ}_-\}=-{1\over 4}\partial_-\lambda_-$
is a consequence of a gauge-invariant definition like (\ref{gaugeinv})
plus one-loop effects.

\appendix{Parity Invariance of (Gauged) WZW Models}
\label{app:D}

In this appendix we discuss the definition of world-sheet parity for
(gauged) WZW models on a group manifold $G$. The WZW action is given by
\beq
S_{\rm WZW}(g)={1\over 8\pi}\int_{\Sigma}
{\rm Tr}\left[(g^{-1}\partial_0g)^2-(g^{-1}\partial_1g)^2\right]
\dd^2x
+{1\over 12\pi}\int_B {\rm Tr}\left[(g^{-1}\dd g)^3\right],
\eeq
where $B$ is a three-dimensional manifold bounded by the two-dimensional
world-sheet $\Sigma$ over which the field $g$ is extended. 
The WZ term depends on the orientation
and is flipped under parity. This can be compensated by
the transformation $g\to g^{-1}$, since
$g^{-1}\dd g\to g\dd g^{-1}=-g(g^{-1}\dd g)g^{-1}$.
The kinetic term is invariant under both parity and $g\to g^{-1}$.
Thus, the WZW model is parity-invariant if accompanied by $g\to g^{-1}$.

Gauging by $g\to h^{-1}gh$ for $h$ in a subgroup $H\subset G$
leads to a vector gauged WZW model, where $g^{-1}\partial_{\mu}g$
in the kinetic term is replaced by $g^{-1}D^{\rm v}_{\mu}g
=g^{-1}\partial_{\mu}g+g^{-1}A_{\mu}g-A_{\mu}$ and the WZ term is modified
by adding
\beq
\Gamma^{\rm v}(A,g)=-{1\over 4\pi}\int_{\Sigma}{\rm Tr}\left[
A(g^{-1}\dd g+\dd g g^{-1})+Ag^{-1}Ag\right].
\label{GvAg}
\eeq
Under $g\to g^{-1}$ the covariant derivative
transforms as $g^{-1}D^{\rm v}g\to -g(g^{-1}D^{\rm v}g)g^{-1}$
and thus the kinetic term is invariant.
Furthermore, it is easy to see that
(\ref{GvAg}) flips sign under this transformation, 
$\Gamma^{\rm v}(A,g^{-1})=-\Gamma^{\rm v}(A,g)$.
Thus, the vector gauged WZW model is parity invariant,
again if accompaied by $g\to g^{-1}$.

Gauging by $g\to h^{-1}gh^{-1}$ for $h$ in
an abelian subgroup $H\subset G$
is another possibility called axial gauging.
The kinetic term is obtained by replacing 
$g^{-1}\partial_{\mu}g\to g^{-1}D^{\rm a}_{\mu}g
=g^{-1}\partial_{\mu}g+g^{-1}A_{\mu}g+A_{\mu},$
and the WZ term is modified by
\beq
\Gamma^{\rm a}(A,g)=-{1\over 4\pi}\int_{\Sigma}{\rm Tr}\left[
A(g^{-1}\dd g-\dd g g^{-1})-Ag^{-1}Ag\right].
\label{GaAg}
\eeq
Under $g\to g^{-1}$ the covariant derivative
transforms as $g^{-}D^{\rm a}g\to
-g(g^{-1}\dd g-g^{-1}Ag-A)g^{-1},$
and thus the kinetic term is invariant only if the sign of $A$ is flipped.
Also, it is straightforward to see that
(\ref{GaAg}) flips sign under $g\to g^{-1}, A\to -A$.
Thus the axially gauged WZW model is parity-invariant
if accompanied by $g\to g^{-1}$ and $A\to -A$.

\subsection*{\it Axially gauged $SL(2,\RR)/U(1)$}

The Euclidean (bosonic or fermionic) 2d Black Hole
is associated with the axial gauging of $SL(2,\RR)$ by the
$U(1)$ generated by $i\sigma_2$.
Thus the parity should act on the fields as
$g\to g^{-1}$ and $A\to -A$.
Setting
$g=\e^{i\sigma_2(t+\varphi)/2}\e^{\rho\sigma_3}\e^{i\sigma_2(t-\varphi)/2},$
we see that the transformation $g\to g^{-1}$ corresponds to
$\rho\to -\rho,
\varphi\to \varphi,
t\to -t$.
(The last one is compatible with $A\to -A$.)
The sign flip of $\rho$ can actually be undone by a $\pi$-shift of $\varphi$.
Hence world-sheet parity can be defined to act on the coordinates as
\beq
\rho\to \rho,~~
\varphi\to\varphi,~~
t\to -t,~~
A\to -A.
\eeq
Next let us describe the action of parity on the current algebra.
Left and right current algebras are associated with the transformation
of the group elements of the form $g\to g_Lgg_R$.
Under $g\to g^{-1}$ this becomes 
$g\to g_R^{-1}g g_R^{-1}$. Thus the right-moving
currents $J^+,J^3,J^-$  are transformed to the left-moving currents
$\tJ^-,\tJ^3,\tJ^+$ and vice versa.
In particular, the right-moving lowest-weight representarions are
transformed to the left-moving highest-weight representations. For
example, the representation $\wh{\cD}_j^{+}\times \wh{\cD}_j^{+}$ is exchanged
with $\wh{\cD}_j^{-}\times \wh{\cD}_j^{-}$.
\footnote{$g\to g_Lgg_R$ is the convention used in \cite{MO}
which leads to the spectrum $\wh{\cD}_j^{\pm}\times \wh{\cD}_j^{\pm}$.
If the other convention $g\to g_Lgg_R^{-1}$ were taken, the spectrum
would be $\wh{\cD}_j^{\pm}\times \wh{\cD}_j^{\mp},$ and
the parity would act as the exchange of
$\wh{\cD}_j^{+}\times \wh{\cD}_j^{-}$ 
and $\wh{\cD}_j^{-}\times \wh{\cD}_j^{+}$.
We thank J. Maldacena for a discussion on this.}

\section*{Acknowledgement}

We would like to thank M.~Aganagic, M.R.~Douglas, P.~Fendley,
K.~Intriligator, A.~Karch, J.~Maldacena, J.~Polchinski,
N.~Seiberg, S.~Shenker, E.~Silverstein, D.~Tong,
C.~Vafa, E.~Witten, and A.B.~Zamolodchikov for discussions.

We thank ITP, Santa Barbara, for the support
under the grant NSF-PHY-9907949.
K.H. would also like to thank Rutgers Physics Department and
the Institute for Advanced Study, for hospitality.
K.H. was supported in part by NSF-DMS 9709694.
A.K. was supported in part by DOE grant DE-FG02-90ER40542.

\end{document}